\documentstyle[pre,amssymb,amsfonts,amsmath,aps]{revtex}
\newcommand{\avl}{\langle\!\langle}
\newcommand{\avr}{\rangle\!\rangle}

\begin{document}

\draft

\title{The Cavity Approach to Noisy Learning in Nonlinear Perceptrons}
\author{Peixun Luo and K. Y. Michael Wong}
\address{
Department of  Physics, Hong Kong University of 
Science and Technology,\\ Clear Water Bay, Kowloon, Hong Kong.   
} 

\date{\today}

\maketitle
  
\begin{abstract}

We analyze the learning of noisy teacher-generated examples
by nonlinear and differentiable student perceptrons using the 
cavity method. The generic activation of an example is a 
function of the cavity activation of the example, which is its 
activation in the perceptron that learns without the example. 
Mean field equations for the macroscopic parameters and the stability 
condition yield results consistent with the replica method. 
When a single value of the cavity activation 
maps to multiple values of the generic activation, there is a competition 
in learning strategy between preferentially learning an example and 
sacrificing it in favor of the background adjustment. We find parameter 
regimes in which examples are learned preferentially or sacrificially, 
leading to a gap in the activation distribution. Full phase diagrams 
of this complex system are presented, and the theory predicts the
existence of a phase transition from poor to good generalization states
in the system. Simulation results confirm the theoretical predictions.    

\end{abstract}


\section{Introduction}
\label{intro}

Since Hopfield's pioneer work on neural networks \cite{Hopfield}, statistical 
mechanics has proved to be a  powerful tool in the study of information
processing. Mean field theories such as the replica method 
\cite{MPV,Seung,Watkin}
and the cavity method \cite{MPV,Mezard,EPL,Opper-win} are successfully 
developed to study these 
problems. In particular, it provides valuable insights to the learning of
examples in neural networks by considering it as an energy minimization
process. Early work used the replica method to study the learning problem in 
various situations \cite{Gardner,Gyorgyi,Amit,Wong-sh}. It has the 
advantage of a readily-used mathematical formalism applicable to general 
cases, and has been applied to linear networks
\cite{Opper-ki,Bos-ki,Bos,Dunmur} and networks with binary outputs   
\cite{Wong-sh,Gardner-de,Bouten-sc,Bouten,Majer,Whyte-sh-w,Whyte-sh},
dealing with learning tasks which are either realizable or non-realizable,
random or teacher-generated data, and clean or noise corrupted data. These 
studies mainly focused on the global properties of the learning system, with 
less emphasis on the microscopic description of the examples and the weights 
in the system. Furthermore, most of these models were still remote from the 
differentiable nonlinear perceptron which is most commonly used today. Other 
work used the Green's function approach which is particularly convenient for 
linear networks \cite{Krogh}, but these systems may not have the 
competitive effects among examples in nonlinear networks, which will be
investigated in this paper. The annealed approximation is suitable for 
analyzing high temperature learning \cite{Seung}, but the results cannot
be directly extended to the more common case of low temperature.

A common phenomenon observed in the studies of learning from examples is the 
existence of phase transitions with abrupt improvement in the generalization 
ability of the networks
once the training examples are sufficiently numerous, or the global 
parameters (e.g.~the weight decay) is suitably tuned \cite{Schwarze-he,%
Schottky,Ahr-bi-s,Kinzel,West-sa}. These transitions are often discontinuous. 
They arises when metastable states are present in the system, leading to 
discontinuous jumps in the network states, hystereses, and the disappearance 
of metastability at spinodal points. Multilayer perceptrons will exhibit a 
transition from permutation symmetric to specialized states 
\cite{Watkin}. In the present paper, we will see these effects in nonlinear 
perceptrons learning noisy examples. Here the competition between the locally 
stable states comes from the different learning strategies used to attain the 
systemwide energy minimum. 

The cavity method is a suitable tool to study information competition effects 
in rule extraction from noisy examples. Large scale neural networks with many 
nodes can be considered as mean field systems since, as far as the learning of 
one example is concerned, the influence of the rest of the examples can be 
regarded as a background satisfying some average properties. The success of the 
mean field approach is illustrated by the capability of the replica method in 
describing the macroscopic properties of neural network learning\cite{Watkin}. 
However, the replica method provides much less interpretation on the processing 
of individual examples since its starting point is the quenched average of the
free energy over the example distribution. The cavity method is an alternative
version of mean field theory. It is a generalization of the 
Thouless-Anderson-Palmer (TAP) approach to spin glasses and starts from a 
microscopic description of the system elements \cite{TAP,MPV}. In this method 
mean-field equations are derived from self-consistent considerations. The 
method was subsequently generalized to learning problems \cite{Mezard,EPL,NIPS} 
and yields macroscopic properties identical to the replica method while at the 
same time provides physical insights to the learning of individual examples. 
Recently, the cavity method was also applied to a number of problems in 
information processing \cite{advanced}.
     
In this paper, we study the learning of noisy examples in nonlinear perceptrons
using the cavity method. Nonlinear networks have the following advantages: 
(i) compared with networks with binary output, gradient descent learning is 
possible, 
(ii) nonlinearity is representative of more complex networks, 
(iii) they have more resemblance with biological neurons \cite{Hertz}. 
Compared with previous studies, we will focus on the effects of information 
competition in the system, and their consequences on the energy landscape, the 
appearance of band gaps in the activation distribution, the choice between 
preferential and even-handed learning strategies as well as their possible 
relationship with phase transitions in this complex system. We analyze 
the parameter regimes with band gaps in the activation distribution, as well as 
the stability condition of the perturbative cavity approach. Simulation results 
show that the assumption of a smooth energy landscape usually works well when 
no gaps are present, but tends to fail when gaps appear. Like spin glass models,
the picture of a smooth energy landscape has to be replaced by one with many 
metastable states that can not be related by perturbative analyses. The 
assumptions of smooth and rough energy landscapes are equivalent to the the 
replica symmetric (RS) and replica symmetry-breaking (RSB) approximations in 
the replica method. The phase diagram of this complex system is shown and the 
occurrence of phase transitions is investigated and compared with simulations.

The rest of this paper is organized as follows. After describing the model in 
the next section, we describe in Sec.~\ref{cavity} the cavity approach and 
introduce the cavity activation, which is the core microscopic variable in the
cavity method. Three self-consistent equations are derived when a smooth energy 
landscape is assumed. In Sec.~\ref{bandgap}, we discuss the case when band gaps 
appear in the activation distribution. Phase transitions in nonlinear 
perceptrons and phase diagrams are the themes of Sec.~\ref{phase}. In 
Sec.~\ref{con} we summarize the results and their implications. Mathmetical 
details are appended at the end of the paper.

\section{THE MODEL}
\label{model}

Consider a student perceptron with $N$ weights $J_j$, $j=1,\ldots,N$ that 
connect the $N$ input nodes and the output node. It is trained to extract the 
rule of a teacher perceptron with the same architecture with $N$ weights $B_j$, 
$j=1,\ldots,N$, where $\langle B_j\rangle=0$ and $\langle B_j^2\rangle=1$. 
A training set of $p$ examples generated by the teacher and corrupted by noise 
is what the student can explore. Each example, 
labeled $\mu$ with $\mu=1,\ldots,p$, consists of an input vector 
${\bbox\xi}^\mu$ and the noisy output $O_\mu$ of the teacher. The input 
components $\xi_j^\mu$ are Gaussian random variables, with 
$\langle\xi_j^\mu\rangle=0$  and 
$\langle\xi_j^\mu\xi_k^\nu\rangle=\delta_{jk}\delta_{\mu\nu}$. The 
activation functions $f(x)$ of both perceptrons are differentiable and 
nonlinear, such as ${\rm sig}(x)\equiv (1-\tanh x)/2$, i.e., the 
teacher and student outputs are respectively  
 \begin{equation} 
 O_\mu\equiv f({\tilde y_\mu})\equiv f(y_\mu+T\eta_\mu)  
 \end{equation}
and
 \begin{equation} 
 f_\mu\equiv f(x_\mu),
 \end{equation}
where $y_\mu\equiv\bbox{B}\cdot\bbox{\xi}^\mu/\sqrt{N}$ 
is the teacher activation, $\eta_\mu$ is Gaussian noise with 
$\langle\eta_\mu\rangle=0$ and $\langle\eta_\mu^2\rangle=1$, $T$ 
is the noise temperature, and 
$x_\mu\equiv\bbox{J}\cdot\bbox{\xi}^\mu/\sqrt{N}$ 
is the student activation. 

During the training procedure, one adapts the student network to minimize an 
energy function that measures the difference between the student 
outputs $f_\mu$ and teacher outputs $O_\mu$ for all training examples.
A natural energy function is the total quadratic error of examples in 
training set, $\sum_{\mu=1}^p(O_\mu-f_\mu)^2\equiv p\varepsilon_t^2$, where we 
call $\varepsilon_t$ training error. However, the final target of learning 
is to get a student perceptron that can generalize well to novel examples, 
i.e., to minimize the generalization error 
$\varepsilon_g\equiv\langle\,[O(\bbox \xi)-f(\bbox \xi)]^2\rangle^{1/2}$,
where $\langle\,\rangle$ is the average performed over the distribution of 
all inputs and the noise. We add a weight decay term to penalize excessively 
long weight vectors and speed up learning, and use the energy function 
   \begin{equation}					\label{energy}
   E=\frac{1}{2}\sum_\mu(O_\mu-f_\mu)^2+\frac{{\bf\lambda}}{2}\sum_jJ_j^2,
   \end{equation} 
where $\lambda$ is the weight decay strength. Minimizing the above energy 
function by gradient descent, one obtains the the equilibrium state
of the student perceptron given by 
   \begin{equation}
   J_j=\frac{1}{\lambda\sqrt{N}}\sum_\mu(O_\mu-f_\mu)f_\mu'\xi_j^\mu, 
							\label{equil}
   \end{equation} 
where the prime in $f_\mu'$ represents the derivative of $f(x_\mu)$. Here we 
are interested in the dependence of the generalization error $\varepsilon_g$ of 
the student perceptron in its equilibrium state (\ref{equil}) on the 
macroscopic parameters, such as the weight decay strength $\lambda$, the noise 
temperature $T$ and the size of training set $\alpha\equiv p/N$. As in 
perceptrons with linear or discrete activation functions, the generalization 
error is essentially determined by the overlap of the student weight vector 
with the teacher weight vector $R$ and the magnitude of the student weight 
vector $q$, which are defined as
   \begin{eqnarray}			  		  \label{qR}
   q=\langle J_j^2\rangle_j & \quad\mathrm{and}\quad  & 
   R=\langle J_j B_j\rangle_j, 
   \end{eqnarray}
where $\langle\,\rangle_j$ represents averaging over the $N$ weights.

\section{THE CAVITY METHOD}
\label{cavity}

In order to get more microscopic understanding of the mechanism in the
learning of neural networks, we use the cavity method
developed in \cite{EPL,NIPS} to tackle the current problem.
After the student perceptron is trained with $p$ examples, it reaches
its energy ground state $\bbox J$ given by Eq.~(\ref{equil}). Suppose a new
example with input vector $\bbox\xi^0$ is fed to the student
perceptron. The activation of example 0 is now given by 
   \begin{equation}
   t_0\equiv\frac{1}{\sqrt{N}}\bbox{J}\cdot\bbox{\xi}^0,
   \end{equation} 
which is called the cavity activation. Since the student ${\bbox J}$ has 
no information about an example it has never learned, the cavity activation 
$t_0$ is a Gaussian variable for random inputs $\xi_j^0$ when $N\gg 1$. It has 
a mean $\avl t_0\avr=0$ and covariances $\avl t_0^2\avr=q$ and 
$\avl t_0 y_0\avr=R$, where $\avl\,\avr$ denotes the ensemble average. Hence 
the distribution of the cavity field is 
   \begin{eqnarray}						\label{pty}	
   P(t_0|y_0)&=&\frac{\exp\big[-\frac{(t_0-Ry_0)^2}{2(q-R^2)}\big]}
		{\sqrt{2\pi(q-R^2)}}.            
   \end{eqnarray}

Trained with all the $p+1$ examples 
$\{(\bbox{\xi}_\mu,O_\mu)|\mu=0,1,\ldots,p\}$,
the student perceptron reaches its equilibrium state ${\bbox J}^0$, with
   \begin{equation}					\label{equil_0}
   J_j^0=\frac{1}{\lambda\sqrt{N}}(O_0-f_0^0)(f_0^0)'\xi_j^0
   +\frac{1}{\lambda\sqrt{N}}\sum_\mu(O_\mu-f_\mu^0)(f_\mu^0)'\xi_j^\mu. 
   \end{equation}    
Here and below, variables with superscript 0 refer to 
those associated with the perceptron ${\bbox J^0}$, which includes example 0 in 
its training set. We see that the generic student activation of example 0, 
$x_0\equiv{\bbox J}^0 \cdot\bbox{\xi}^0/\sqrt{N}$, is no longer a 
Gaussian variable. (Although the correct notation of $x_0$ should be
$x_0^0$, here we omit the superscript since it is sufficiently distinct 
from its cavity counterpart $t_0$.) However, it is reasonable to assume that 
the difference between ${\bbox J}$ and ${\bbox J^0}$ is small; the validity
of this assumption will be discussed later. Following the perturbative analysis 
in \cite{EPL}, we show in Appendix \ref{app-a} that, for a given corrupted 
teacher output ${\tilde y}_0$, there is a well defined relation between $t_0$ 
and $x_0$, $t_0=t(x_0,{\tilde y}_0)$, where
   \begin{equation}					      \label{tx}
   t(x,{\tilde y})=x-\gamma[f({\tilde y})-f(x)]f'(x).
   \end{equation} 
Here the parameter $\gamma$ is the local 
susceptibility and satisfies 
   \begin{equation}                                            \label{sus}
   1-\gamma\lambda=\alpha\Big\langle 1-\frac{\partial{x_\mu}}
       			{\partial{t_\mu}}\Big\rangle_\mu,
   \end{equation} 
where $x_\mu$ is a single-valued function of $t_\mu$, and 
$\langle\cdot\rangle_\mu$ represents averaging over the $p$ examples. 
In this section we will focus on the case that Eq.~(\ref{tx}) presents a 
one-to-one mapping between $x_\mu$ and $t_\mu$ for a given ${\tilde y}_\mu$. 
As we shall see, this corresponds to a continuous activation distribution
with no band gaps. In the next section we will discuss the case when $t_\mu$
has a one-to-many relation with $x_\mu$, which will lead to the emergence of
band gaps.

Combining Eqs.~(\ref{pty}) and (\ref{tx}), we can derive the student activation 
distribution $P(x|{\tilde y},y)$,
   \begin{equation}					\label{pxyy}
   P(x|{\tilde y},y)=P{\bbox (}t(x,{\tilde y})|y{\bbox )}
       \frac{\partial t(x,{\tilde y})}{\partial x}.
   \end{equation}
In turn, the distributions $P({\tilde y}|y)$ and $P(y)$ are given by
   \begin{eqnarray}			
   P({\tilde y}|y)&=&\frac{1}{\sqrt{2\pi T^2}}\exp
	\big[-\frac{({\tilde y}-y)^2}{2T^2}\big]           \label{pyy}  \\
   P(y)&=&\frac{1}{\sqrt{2\pi}}\exp(-\frac{y^2}{2}).    \label{py}
   \end{eqnarray} 
Equation (\ref{sus}) for $\gamma$ can now be transformed to an integral
expression when $N$ approaches infinity,
   \begin{equation}						\label{susi3}
    1-\gamma\lambda=\alpha\!\int\!\!{\rm d} yP(y)\!\int\!\!{\rm d}
	{\tilde y}P({\tilde y}|y)\!\int\!\!{\rm d} xP(x|{\tilde y},y)
              \big(1-\frac{\partial x}{\partial t}\big),
   \end{equation}  
where $\partial x/\partial t={\bbox (}1+\gamma\{[f'(x)]^2-[f({\tilde y})-f(x)]
f''(x)\}{\bbox )}^{-1}$. Equation (\ref{susi3}) can be simplified into an 
equation involving only double integrals,
   \begin{equation}						\label{susi2}
    1-\gamma\lambda=\alpha\int{\rm D}u\int{\rm D}v{\bbox [}1-{\bbox (}
       1+\gamma\{[f'(x)]^2-[f({\tilde y})-f(x)]f''(x)\}{\bbox )}^{-1}{\bbox ]}
   \end{equation}  
where ${\rm D}u\equiv{\rm d}u\,\exp(-u^2/2)/\sqrt{2\pi}$ and 
${\rm D}v\equiv{\rm d}v\,\exp(-v^2/2)/\sqrt{2\pi}$ are Gaussian measures, 
${\tilde y}=\sqrt{1+T^2}u$, and $x$ depends on $u$ and $v$ via 
   \begin{equation}						\label{xuv}	
   \frac{R}{\sqrt{1+T^2}}u+\sqrt{q-\frac{R^2}{1+T^2}}\,v
	=t(x,{\tilde y}).
   \end{equation} 

The mean field equation for $R$ can be obtained by multiplying both sides of 
Eq.~(\ref{equil}) with $B_j$, and summing over $j$, yielding 
     \begin{eqnarray}
     R=\frac{\alpha}{\lambda}\int{\mathrm d}yP(y)\int{\mathrm d}{\tilde y}
                P({\tilde y}|y)\int{\mathrm d}x
		P(x|{\tilde y},y)[f({\tilde y})-f(x)]f'(x)y. 
     \end{eqnarray}
Substituting Eqs.~(\ref{pxyy}-\ref{py}), we can simplify it to 
     \begin{equation}
     R=\frac{\alpha}{\lambda}\int{\mathrm D}u\int{\mathrm D}v[f(\sqrt{1+T^2}u)
        -f(x)]f'(x)\Big[\frac{1}{\sqrt{1+T^2}}u+\frac{RT^2}{(1+T^2)\sqrt{q-
        \frac{R^2}{1+T^2}}}v\Big],
     \end{equation}
where $x$ depends on $u$ and $v$ via Eq.~(\ref{xuv}).
Integrating by parts and using Eq.~(\ref{susi2}) and (\ref{xuv}), we have
    \begin{eqnarray}      		
    R=\alpha\gamma\int{\mathrm D}u\int{\mathrm D}v\frac{f'({\tilde y})
	f'(x)}{1+\gamma\{[f'(x)]^2-[f({\tilde y})-f(x)]f''(x)\}}.   \label{R}
    \end{eqnarray}
Similarly, multiplying both sides of Eq.~(\ref{equil}) by $J_j$, and summing 
over $j$, we have
    \begin{equation}						
    q=\frac{\alpha}{\lambda}\int{\mathrm d}yP(y)\int{\mathrm d}{\tilde y}
                P({\tilde y}|y)\int{\mathrm d}x
		P(x|{\tilde y},y)[f({\tilde y})-f(x)]f'(x)x.
    \end{equation} 
Again, simplifying into double integrals and integrating by parts, we arrive at 
    \begin{equation} 
    q-R^2=\alpha\gamma^2\int{\mathrm D}u\int{\mathrm D}v
	  [f({\tilde y})-f(x)]^2[f'(x)]^2.      \label{q} 
    \end{equation} 

The three macroscopic parameters $\gamma$, $R$ and $q$ can now be obtained by 
solving the three mean field equations (\ref{susi2}), (\ref{R}) and (\ref{q}) 
numerically for given values of $\alpha$, $\lambda$ and $T$. Therefore we can 
directly obtain the training error $\varepsilon_t$ and generalization error 
$\varepsilon_g$, which depend on the generic activation $x$ and cavity 
activation $t$ respectively, 
  \begin{eqnarray}   
  \varepsilon_t^2&=&\int{\mathrm D}u\int{\mathrm D}v[f({\tilde y})-f(x)]^2, \\
  \varepsilon_g^2&=&\int{\mathrm D}u\int{\mathrm D}v[f({\tilde y})-f(t)]^2.	
  \end{eqnarray}

The validity of the perturbative calculation can be checked by considering
the stability condition of the equilibrium state. As derived in Appendix 
\ref{app-b}, when the new example 0 is added, the magnitude of the change in 
the student weight vector is given by
   \begin{equation}						\label{Delta_J}
  \Delta_J\equiv\sum_j(J_j^0-J_j)^2=\frac{(x_0-t_0)^2}{1-\alpha\Big\langle\big
   (1-\frac{\partial x_\mu}{\partial t_\mu}\big)^2\Big\rangle_\mu}.
   \end{equation}
Hence $\Delta$ diverges when the denominator approaches 0. This yields the 
stability condition   
   \begin{equation}						\label{stab}
   \alpha\Big\langle\big(1-\frac{\partial x_\mu}{\partial t_\mu}
     \big)^2\Big\rangle_\mu<1. 
   \end{equation} 
It is identical to the stability condition of the replica-symmetric (RS)
ansatz in the replica approach \cite{Wong-sh,EPL}, the so-called 
Almeida-Thouless (AT) condition \cite{Almeida-t}. 

In the region where the stability condition (\ref{stab}) is violated,
the perturbative version of cavity method breaks down. It becomes possible
that when a new example is added to the system, the ground state relocates 
to another metastable state. This corresponds to the picture of a rough energy
landscape with many metastable states and the perturbative cavity method has 
to be modified \cite{NIPS}.

\section{ACTIVATION DISTRIBUTIONS WITH BAND GAPS}
\label{bandgap}

When the activation function $f(x)$ is nonlinear, the behavior of the 
system may be very complex. This can be seen by considering Eq.~(\ref{tx})
for a sufficiently large $\gamma$, when the generic activation $x$ may 
become a multi-valued function of the cavity activation $t$. In such a
case, the system settles at its ground state, i.e., chooses the value 
of $x$ that minimizes the energy function in Eq.~(\ref{energy}). 

The energy increase on adding example 0 can be derived easily. According to 
Eq.~(\ref{energy}), the energy difference between the perceptron states 
$\bbox{J}^0$ and $\bbox{J}$ is
   \begin{equation}
   \Delta E\equiv E^0-E=\frac{1}{2}(O_0-f_0)^2+\frac{1}{2}\sum_\mu[(O_\mu-f_
         \mu^0)^2-(O_\mu-f_\mu)^2]+\frac{\lambda}{2}\sum_j[(J_j^0)^2-J_j^2].  
   \end{equation}     
Expanding the first summation to the second order $(x_\mu^0-x_\mu)^2$ and 
substituting Eq.~(\ref{equil}) and Eq.~(\ref{equil_0}) to the second 
summation, we can simplify the above equation to 
   \begin{equation}
   \Delta E=\frac{1}{2}(O_0-f_0)^2+\frac{1}{2}(O_0-f_0)f_0'(x_0-t_0).
   \end{equation}
Using the relation between the cavity activation $t_0$ and generic activation 
$x_0$ in Eq.~(\ref{tx}), we find
   \begin{equation}					\label{change}
   \Delta E=\frac{1}{2}(O_0-f_0)^2+\frac{1}{2\gamma}(x_0-t_0)^2. 
   \end{equation}
The first term is the primary change due to the newly added example,
and the second term results from the adjustment of the background examples. 
In the multi-valued region, one needs to compare the energy increase of the 
solutions whose values of $x_0$ are closer to $t_0$ (therefore favorable to 
small background adjustment) with those whose outputs $f_0$ are closer to the 
teacher's outputs $O_0$ (therefore favorable to small primary cost). This 
competition leads to a discontinuity in the range of the generic activation 
$x_0$ when the cavity activation $t_0$ varies, accompanied by the appearance 
of gaps in the activation distribution for a given teacher output.

To study this competition, we suppose that Eq.~(\ref{tx}) has multiple 
solutions of $x$ in a range of $t$, for a given ${\tilde y}$. We are interested 
in the point $t_g({\tilde y})$ where two solutions yield the same energy 
change $\Delta E$. That is, there are two distinct values of $x$, $x_<$ and 
$x_>$, such that $t_g=t(x_<,{\tilde y})=t(x_>,{\tilde y})$ and $\Delta E(x_<)
=\Delta E(x_>)$. Then using Eq.~(\ref{change}), we arrive at the condition 
   \begin{equation}
   \int_{x_<(t_g)}^{x_>(t_g)}t(x){\rm d} x=t_g[x_>(t_g)-x_<(t_g)],
   \end{equation}
which is the Maxwell's construction as shown in Fig.~\ref{Fig_x-t}. As the 
result of energy minimization, one of the two solutions of $x$ is preferred on 
the left neighborhood of $t_g$, while the other is preferred on the right. 
Hence $x$ is a function of $t$ with a discontinuity at $t_g$. 

Consequently, the student activation distribution $P(x|{\tilde y})$ for a given
teacher output ${\tilde y}$ becomes zero when the student activation $x$ is 
located in the band gap $[x_<(t_g),x_>(t_g)]$. Due to the effect of band gaps 
in $P(x|{\tilde y})$, extra terms should be added to the mean field 
equations Eq.~(\ref{susi2}) and Eq.~(\ref{R}) for $\gamma$ and $R$, as derived 
in Appendix C, namely, 
    \begin{eqnarray}						
    1-\gamma\lambda&=&\alpha\int\{{\rm D} u\sum_i\int_{R_i}{\rm D} v
        {\bbox [}1-{\bbox (}1+\gamma\{[f'(x)]^2-[f({\tilde y})-f(x)]f''(x)\}
         {\bbox )}^{-1}{\bbox ]}                        	  \nonumber\\
      &&-\alpha\int{\mathrm D}u\sum_j G(t_g^j)[x_>(t_g^j)-x_<(t_g^j)],
							        \label{sus_e}\\
    R&=&\alpha\gamma\int{\rm D} u\sum_i\int_{R_i}{\rm D} v\frac{f'({\tilde y})
      f'(x)}{1+\gamma\{[f'(x)]^2-[f({\tilde y})-f(x)]f''(x)\}}   \nonumber \\
      &&+\alpha\gamma\int{\rm D} u\sum_j G(t_g^j)f'({\tilde y}){\bbox [}f{\bbox
       (}x_>(t_g^j){\bbox )}-f{\bbox (}x_<(t_g^j){\bbox )}{\bbox ]},\label{R_e} 
     \end{eqnarray}  
where each term in the summations over $i$ corresponds to an integration over
a region $R_i$ separated from each other by band gaps, and each term in the  
summation over $j$ corresponds to a band gap. The Gaussian factor $G(t_g^j)$
is given by 
     \begin{equation}					       \label{ptg}
     G(t_g^j)=\frac{1}{\sqrt{2\pi\big(q-\frac{R^2}{1+T^2}\big)}}
      \exp\Big[-\frac{\big(t_g^j-\frac{R}{\sqrt{1+T^2}}u\big)^2}
      {2\big(q-\frac{R^2}{1+T^2}\big)}\Big].
     \end{equation}        
We note that the extra terms due to gaps are consistent with adding the delta
function component $(x_>-x_<)\delta(t-t_g)$ to $\partial x/\partial t$ in 
Eq.~(\ref{susi2}) and $[f(x_>)-f(x_<)]\delta(t-t_g)$ 
to $f'(x)\partial x/\partial t$ in Eq.~(\ref{R}).

For the sigmoid function $f(x)=(1+e^{-x})^{-1}$, the necessary and sufficient
condition for Maxwell's construction, as derived on Appendix D, is
   \begin{equation}				\label{W_ga}
   \bigg\{ \begin{array}{lll}
   f({\tilde y})-\frac{1}{2}>W(\gamma) \quad & {\rm for}  & \quad x<0   \\
   \frac{1}{2}-f({\tilde y})>W(\gamma) \quad & {\rm for}  & \quad x>0,
            \end{array}             
   \end{equation}
where the function $W(\gamma)$ is monotonic, as shown in 
Fig.~\ref{Fig_W-gamma}. The behavior of the activation distribution depends 
on the value of $\gamma$ in the following three cases:

{\it Case 1:} $\gamma<(117+165\sqrt{33})/64\approx 16.64$. As $W(\gamma)>1/2$ 
and $0\le f({\tilde y})\le 1$, the condition (\ref{W_ga}) cannot be satisfied 
for all teacher output $f({\tilde y})$. Hence there is no gap in the activation 
distribution.

{\it Case 2:} $16.64<\gamma<48$. Here $0<W(\gamma)<1/2$. The 
activation distribution starts to develop a band gap which 
extends from $f({\tilde y})=1$ to $f({\tilde y})=1/2+W(\gamma)$ in the region 
$f(x)<1/2$. Similarly, another band gap extends from $f({\tilde y})=0$ to 
$f({\tilde y})=1/2-W(\gamma)$ in the region $f(x)>1/2$. The two band gaps are
symmetric with respect to the point 
${\bbox (}f(x),f({\tilde y}){\bbox )}=(1/2,1/2)$. For 
intermediate teacher output between $1/2\pm W(\gamma)$, the distribution
remains continuous. The gapped regions are bounded by solid lines in 
Fig.~\ref{Fig_Maxwell}.

{\it Case 3:} $\gamma>48$. Here $W(\gamma)<0$. The band gap in the region 
$f(x)<1/2$ now extends from $f({\tilde y})=1$ to $f({\tilde y})=1/2+W(\gamma)
<1/2$. Together with its symmetric counterpart in the region $f(x)>1/2$, the 
activation distribution is three-banded for $1/2+W(\gamma)<f({\tilde y})<1/2
-W(\gamma)$, beyond which activation distribution remains two-banded. 

When a band gap is present for a given $f({\tilde y})$, multiple energy 
minima can exist for a generic student activation $x$ near the gap.
When the energy minimum favors the generic activation to take a
value closer to the teacher activation than the cavity activation, the example 
is {\it preferentially} learned. Otherwise, when the generic activation is 
closer to the cavity activation, the example is {\it sacrificed}.  
As shown in Fig.~\ref{Fig_Maxwell}, a band gap exists in the regions that are 
shaded or enclosed by the transition lines 
$L_s^{tr}$ and $L_p^{tr}$ (subscripts {\it s} and {\it p} represent sacrificed 
and preferred states respectively). In the neighborhood of the sacrificed band 
edge, the line $L_s^c$ indicates the onset of competition. Between the lines 
$L_s^c$ and $L_s^{tr}$, the sacrificed state is competing with a metastable 
preferred state, which appears between the spinodal line $L_p^{sp}$ and the 
line $L_p^{tr}$. As illustrated in Fig.~\ref{Fig_x-t}, the stable states 
between points $P_<^c$ and $P_<^{tr}$ competes with metastable states between 
points $P_>^{sp}$ and $P_>^{tr}$, but the sacrificed states 
remain the ground states. Between line $L_s^{tr}$ and the spinodal line 
$L_s^{sp}$, the sacrificed state becomes metastable. It disappears at 
$L_s^{sp}$. Similar lines exist in the neighborhood of the preferred band 
edge.
 
The condition of the band gaps and preferential learning shown in 
Fig.~\ref{Fig_Maxwell} is for $\gamma=20.55$. In the ground state, no examples 
exist in the shaded region, rendering a gap of the activation distribution. One 
finds that preferential learning first occurs at extreme values of the 
teacher output, 
$f({\tilde y})>f({\tilde y^+})=0.864$ or $f({\tilde y})<f({\tilde y^-})=0.136$. 
For $f({\tilde y})<0.136$, student activations to the left 
of the shaded region correspond to the preferred examples, whereas those to the 
right correspond to the sacrificed ones. 
The energy advantage of this learning strategy can be easily understood. In
nonlinear perceptrons, changes in the student activation around these extreme 
values of $f({\tilde y})$ do not result in significant changes in the training 
error of an example due to the saturation in this region, and if the cavity 
activation is very different from the teacher's activation, it 
is more economical to keep the student activation close to 
the cavity activation, so that the background adjustment remains small. In 
contrast, for intermediate values of $f({\tilde y})$, the competitive effects 
are less, and no band gaps develop. 

The width of the band gap can be narrowed when the existence of metastable 
states is taken into account. As shown in Fig.~\ref{Fig_Maxwell}, metastable 
states exist inside the band gap as far as the spinodal lines $L_s^{sp}$ and 
$L_p^{sp}$. Hence in finite-time simulations, the system may be trapped in 
metastable states. Conventionally, the narrowing of band gaps in simulations is 
explained by RSB effects in the replica method \cite{Majer,Whyte-sh-w}. Here we 
conclude that the narrowing can be explained by metastability in the 
perturbative cavity method, which is equivalent to the replica symmetric 
ansatz, without invoking the formalism of RSB.

For comparison, this kind of preferential learning is not present in linear 
perceptrons, even when perfect learning is impossible. Since $f(x)=k(x+\theta)$,
the activation $x$ becomes a linear function of the cavity activation $t$, by 
virtue of Eq.~(\ref{tx}). Hence preferential learning is a unique 
consequence of the nonlinearity of the perceptron activation. 

Figure \ref{Fig_gap-regime} shows the parametric regimes for the existence of 
gapped activation distributions as well as the unstable regimes of the 
perturbative cavity method (the boundary line being equivalent to the AT line 
in the replica method) for different noise temperature. Since the development 
of a gap is already sufficient to cause an uncontrollable change in 
(\ref{Delta_J}), 
the gapped regions lie inside the unstable regions. Furthermore, provided that 
$\alpha$ and $T$ are not too large, the boundaries of the gapped and unstable 
regions are very close to each other. The region of small weight decay and 
large noise will be discussed in the next section, where the phase lines are 
modified when discontinuous transitions take place.  

It is shown in Fig.~\ref{Fig_gap-regime} that band gaps in the activation 
distribution exist when 
the training set size $\alpha$ is small, noise temperature $T$ is large and 
weight decay strength $\lambda$ is insufficient, leading to the preferential 
learning of some examples while sacrificing others. When the training examples 
are sufficient, the underlying rule can be extracted with confidence, thereby 
restoring the continuous distribution. Furthermore, increasing the data noise 
broadens the gapped region. Indeed, noisy data introduces conflicting 
information to be learned by the student. On the other hand, the gapped region 
narrows with increasing weight decay strength. Arguably, weight decay restricts 
the flexibility in the weight space, thus reducing the tendency for multiple 
minima.

We check the appearance of band gaps predicted in our theory with simulations. 
Four typical activation distributions, all with $\alpha=3$ and $\lambda=0.002$, 
are plotted in Figs.~\ref{Fig_d-a3wd0002}(a-d) at increasing $T$. To facilitate 
comparison, examples are collected for the noise-corrupted teacher activation 
${\tilde y}$ at the value of $-\sqrt{1+T^2}$, so that the probabilities 
$P(\tilde{y})$ are the same.

Figure \ref{Fig_d-a3wd0002}(a) is the distribution at $T=0.1$, where 
$\gamma=11.1$ and the stability condition (\ref{stab}) is fulfilled. The 
student activation distribution in this case has a single band and is a sharp
peak at $x={\tilde y}$. When noise temperature $T$ 
increases to 2 where $\gamma=14.9$, the location of the parameters $\alpha=3$ 
and $\lambda=0.002$ in Fig.~\ref{Fig_gap-regime} is slightly above the boundary 
between gapped and continuous regimes. Correspondingly, there is a pseudogap 
developed in the activation distribution, as shown in 
Fig.~\ref{Fig_d-a3wd0002}(b). Comparing with simulation 
results, we see that the assumption of a smooth energy landscape used in the 
present work is valid in this regime. As shown in Table I, the theoretical and 
simulation results of the training error $\varepsilon_t$, the generalization 
error $\varepsilon_g$, the weight overlap $R$ between teacher and student and 
the student weight magnitude $q$ also agree well.

When $T=2.5$, where $\gamma=20.6$, the stability condition is 
violated and there is now a gap in both the prediction of the cavity method 
and the simulation result, as shown in Fig.~\ref{Fig_d-a3wd0002}(c). However, 
we can see from Fig.~\ref{Fig_d-a3wd0002}(d), where $T=5$, the theoretical 
prediction of the band gap is broader and has sharper edges than the 
simulation one. At the same time, as shown in Table I, there are prominent 
differences of $\varepsilon_t$, $R$ and, especially, $q$. Two arguments are 
relevant. First, the narrowing of the band gap can be explained by the presence 
of metastable states in the band gap as discussed in Fig.~\ref{Fig_Maxwell}. 
These metastable states probably prevent the learning process to converge to 
the ground state, which therefore yields a value of $q$ different from the 
theory. Secondly, due 
to the violation of the stability condition (\ref{stab}) when the band gap 
develops, a rough energy landscape as discussed previously \cite{NIPS} must be 
introduced to improve the agreement. It is hopeful that the first step 
replica symmetry-breaking ansatz will predict a more consistent result 
with the simulation, such as a shallower band gap and a smaller value of $q$.
For exact solutions, it was pointed out recently that whenever there is gap in 
the activation distribution, full RSB analysis is necessary \cite{Whyte-sh}.

In Fig.~\ref{Fig_d-wd0001t5}, we plot the activation distributions predicted 
theoretically for different $\alpha$ with the same noise temperature $T=5$ and 
weight decay strength $\lambda=0.001$. One finds that while 
Fig.~\ref{Fig_gap-regime} shows that insufficient 
examples cause the appearance of band gaps, here one finds it is possible that
the fraction of examples located in the sacrificed band decreases with the size
of the example set. Therefore, the competitive effects of learning strategies 
are serious only when both noise and the size of training set are large, as one 
may expect intuitively. 

\section{PHASE TRANSITIONS}
\label{phase}

Another consequence of nonlinearity is the existence of two stable solutions of 
$R$, $q$ and $\gamma$ to the mean field equations (\ref{sus_e}), (\ref{R_e}) 
and (\ref{q}) for a given set of parameters. We plot the curves of $\lambda$ 
versus $\gamma$ in Fig.~\ref{Fig_lambda-ga} for different values 
of $\alpha$ at a given noise temperature $T$. Studying the behaviors of the 
curves, and hence the accompanying phase transition in different ranges of 
$\alpha$, we find two critical parameters $\alpha_c^*(T)$ and $\alpha_0(T)$ for 
a given noise temperature $T$. [$\alpha_c^*(1)=1.65$, $\alpha_0(1)=1.737$.] 

{\it Case 1:} $\alpha<\alpha_c^*(T)$. $\lambda$ is a monotonic decreasing 
function of $\gamma$. Hence for any weight decay strength, there is  
a unique local susceptibility. Numerical results in this region shows that the 
magnitude of student weight vector $q$ increases with decreasing weight decay 
$\lambda$.

{\it Case 2:} $\alpha_c^*(T)\le\alpha<\alpha_0(T)$.
At $\alpha=\alpha_c^*(T)$, multiple solutions of $\gamma$ for a given $\lambda$
start to appear near the inflection point of the curve. The solution with the
smallest $\gamma$ corresponds to the {\it good generalization } solution with 
small $q$ and small $\varepsilon_g$. The solution with the largest $\gamma$ 
corresponds to the {\it poor generalization} solution, with large $q$ and 
$\varepsilon_g$. In between the two solutions, there is a third, unstable, 
solution, which can be considered as the barrier separating the two stable 
solutions in the energy landscape. When $\alpha$ increases beyond 
$\alpha_c^*(T)$, the intermediate range of $\lambda$ for which multiple 
solutions exist becomes increasing wide.

At very large $\lambda$, the good generalization state is the only stable 
solution. When $\lambda$ decreases, a metastable state with poor generalization 
appears at the {\it spinodal point} $\lambda_p(\alpha,T)$. When $\lambda$ 
decreases further, the globally stable state switches from the good 
generalization state to the poor as a discontinuous phase transition at 
$\lambda_c(\alpha,T)$. The point where this first-order transition occurs has 
to be determined by comparing the energy of the two states. On further decrease 
of $\lambda$, the metastable state of good generalization disappears at another 
{\it spinodal point} $\lambda_g(\alpha,T)$. Hence $\alpha_c^*(T)$ 
is a {\it critical point} where discontinuous transition first appears. 

{\it Case 3:} $\alpha>\alpha_0(T)$. At $\alpha=\alpha_0(T)$, the {\it 
spinodal point} $\lambda_g$ of the good generalization state vanishes. Hence 
both poor generalization and good generalization solutions coexist for 
$\lambda$ below $\lambda_p$ down to zero. Here the example set is large enough 
to provide information about the teacher such that the good generalization 
solution exists even in the absence of weight decay, although it is only 
metastable.

The dependence of the generalization error $\varepsilon_g$ on the weight decay 
strength $\lambda$ is shown in Fig.~\ref{Fig_energy-eg-t1} for different sizes 
$\alpha$ of training set. When $\alpha<\alpha_c^*(T)$, there is no phase 
transition, and the generalization 
error decreases continuously on increasing $\lambda$ till the optimal weight 
decay strength $\lambda_{\rm opt}(T)(\approx 0.05$, for $T=1$), where the 
perceptron generalizes best. Similar phenomena are also found in linear 
perceptrons learning noisy examples and constrained with weight decay 
\cite{Dunmur}. However, the situation in nonlinear perceptron learning noisy 
examples becomes more complex for larger $\alpha$. When $\alpha>\alpha_c^*(T)$, 
the energy curve has two stable branches that cross at $\lambda_c(\alpha,T)$ 
(as illustrated in the inset of Fig.~\ref{Fig_energy-eg-t1}). Here the 
thermodynamic transition of $\varepsilon_g$ takes place discontinuously. In the 
range of $\lambda$ around $\lambda_c$, metastable states separated by an energy 
barrier exist. Hence in practice, the transition between the good and poor 
generalization states may not take place at the same point, leading to 
hysteretic effects.

The existence of the discontinuous transition when $\lambda$ changes, 
accompanied by the hysteretic effects, is verified by the simulation of a
sample in Fig.~\ref{Fig_hestera4t5}, where $\alpha=4,T=5$. It is interesting to 
observe a third state with intermediate $q$ and $\varepsilon_g$. The existence 
of such intermediate states is not uncommon in simulations, although 
transitions between the poor and good generalization are mostly direct, as 
predicted by the theory. Considering the stability condition (\ref{stab}) for 
the parameters used in Fig.~\ref{Fig_hestera4t5}, we 
find that the perturbative cavity solution is stable in the good generalization 
phase, but unstable in the poor one. This implies that multiple metastable 
states may exist in the poor generalization phase, contributing to the 
cascading transition observed in Fig.~\ref{Fig_hestera4t5}.

Similarly, discontinuous  transitions occur when $\alpha$ increases for a given
$\lambda$. The learning curve for different weight decay strengths is plotted 
in Fig.~\ref{Fig_t1-t5-wd}(a) for $T=1$ and in Fig.~\ref{Fig_t1-t5-wd}(b) for 
$T=5$. One see that in both cases, 
the student may even learn worse for more training examples if the training
examples are not sufficient. Only after sufficient examples are fed to the 
student will $\varepsilon_g$ decrease asymptotically on increasing $\alpha$. 
For smaller weight decay, there is a discontinuous transition from a good 
to a poor generalization state at a critical example size 
$\alpha_c(\lambda,T)$. Discontinuous transitions on changing $\alpha$ and 
$\lambda$ are also observed in the high temperature limit in multilayer 
networks learning clean examples \cite{Ahr-bi-s}.
 
Sample averaged simulations for $T=1$ and $\lambda=0.0001$, as shown 
in Fig.~\ref{Fig_t1-t5-eg}(a), show that theory and simulation agree 
satisfactorily on both sides of the bump. However, theory predicts a relatively 
abrupt change of $\varepsilon_g$ for $\alpha$ around 1.6, which is not observed 
in the simulation. This discrepancy may be partly due to the finite size 
effects, but we cannot preclude that effects of rough energy landscape (RSB) 
also contribute.

This discrepancy between theory and sample averaged simulations is also 
observed at $\lambda=0.001$ and $T=5$, as shown in Fig.~\ref{Fig_t1-t5-eg}(b). 
Here we see that discontinuous transitions exist for large noise temperature 
$T$. Hysteretic effects are shown by the different values of the transition 
points in the upward and downward directions of changing $\alpha$, given by 
$\alpha_c^u(\lambda,T)=4.84$ and $\alpha_c^d(\lambda,T)=4.29$ respectively. The 
theoretical prediction of $\alpha_c(\lambda,T)$ is obtained in 
Fig.~\ref{Fig_t5-e-q-a}  from the 
intersection of the 
energy curves of the branches of poor and good generalization states. However, 
this prediction of $\alpha_c^t(\lambda,T)=5.95$ is higher than
the position of hysteresis. Again, we attribute the discrepancy to finite size 
and the rough energy landscape.

We can interpret the effects of a rough energy landscape from the comparison 
between theoretical and simulation results. As shown in 
Fig.~\ref{Fig_t1-t5-eg}(b), the correction due to a rough energy landscape is 
minor for small $\alpha$. Although a band 
gap exists in the activation distribution, the statistical weight of the 
outlying bands is only very small, as shown in Fig.~\ref{Fig_d-wd0001t5}. When 
the size of the training set increases, the increasing weight of the outlying 
bands, as shown in Fig.~\ref{Fig_d-wd0001t5}, implies stronger effects of rough 
energy landscapes, which may account for the lowering of the critical $\alpha$ 
of the discontinuous transition in simulations when compared with the 
prediction of a smooth energy landscape. The smooth ansatz is stable for the 
branch of good generalization state in Fig.~\ref{Fig_t5-e-q-a}, but unstable 
for the poor one. Hence the introduction of the roughening effects will 
modified the energy curve of the poor generalization state, while that of the 
good one remains unchanged, thus shifting the position of the crossing point. 
The lowering of the $\alpha$ value in simulations implies that the energy of 
the poor generalization state is higher when we change from a smooth picture to 
a rough one. This is consistent with previous results that RSB increases the 
energy of similar perceptrons with discrete outputs \cite{Majer,Whyte-sh}. 
 
The full phase diagram is drawn in Fig.~\ref{Fig_phase-diagram-t1}  for a given 
noise temperature $T$. 
Above and below the {\it thermodynamic transition }line, line $a$, the 
perceptron is in the good and poor generalization phase respectively. Line $a$ 
ends at the critical point $P$, where $\alpha=\alpha_c^*(T)$. The values of $q$ 
and $\varepsilon_g$ change discontinuously when the global parameters move 
cross line $a$, but continuously when they move around point $P$ without 
crossing line $a$. The difference between the continuous and discontinuous 
transitions around point $P$ is illustrated in Fig.~\ref{Fig_t1-t5-eg}(a-b). 
A discontinuous transition across line $a$ below $P$ 
would look like Fig.~\ref{Fig_t1-t5-eg}(b) whereas a bumpy crossover, rather 
than a sharp transition, would take place above $P$ as shown in 
Fig.~\ref{Fig_t1-t5-eg}(a). Cuspy behavior similar to that in 
Fig.~\ref{Fig_t1-t5-eg}(a) is also found in linear networks learning 
un-realizable tasks \cite{Bos-ki,Bos}. Further above $P$, the bump smoothes out 
and the position of $\alpha$ with maximum  $\varepsilon_g$ shifts towards 1, as 
shown in Fig.~\ref{Fig_t1-t5-wd} for increasing $\lambda$. The position of the 
maximum also depends on the noise temperature $T$. For small values of $T$, the
maximum stays near $\alpha=1$, but for larger noise, the maximum could move to 
higher values of $\alpha$, which implies that more examples are required for 
the student to really learn some essence of the teacher's rule when the noise 
is stronger.

Line $b$ denotes the stability line separating the regimes of smooth and rough
energy landscapes. The rough regime covers the entire region left of the 
stability line as well as the entire poor generalization phase below line $a$. 
Here the position of line $a$ is estimated assuming smooth energy landscape. 
Simulations such as those in Fig.~\ref{Fig_t1-t5-eg}(b) indicate that the 
effects of rough energy landscapes may shift its position leftwards. The 
boundary line between gapped and ungapped regions is effectively 
indistinguishable with the stability line at $T=1$. Line $c$ is the 
{\it spinodal} line of the poor generalization phase, where 
$\lambda=\lambda_p(\alpha,T)$. The poor generalization phase is metastable in 
the shaded region bounded by the lines $c$ and $a$. Similarly, line $d$ is the 
{\it spinodal} line of the good generalization phase, where 
$\lambda=\lambda_g(\alpha,T)$, with the good generalization phase 
being metastable between lines $d$ and $a$. When $\lambda$ approaches zero, the 
abscissa of line $d$ approaches $\alpha_0(T)$. Both lines $c$ and $d$ are 
computed in the smooth ansatz only, with roughening effects neglected.

It is interesting to consider the change of learning strategy in different 
regions of Fig.~\ref{Fig_phase-diagram-t1}. In the region bounded by lines $c$ 
and $d$, more than one learning strategies are competing against each other, 
corresponding to different local minima in energy. To the left of line $b$, all 
states adopt learning strategies which sacrifice a fraction of examples, but 
those with large $q$ (poor generalization as shown in Fig.~\ref{Fig_t5-e-q-a})  
sacrifice a significantly large fraction. To the right of line $b$, the 
competition takes places between states with large $q$, which sacrifice a 
fraction of examples, and those with small $q$ (good generalization as shown in 
Fig.~\ref{Fig_t5-e-q-a}), which use a more even-handed strategy with no band 
gaps separating the activations. Around the phase transition line $a$, the 
globally minimal state switches, on increasing $\alpha$, from one with 
sacrificial strategy to a more even-handed one. This discontinuous change in 
learning strategies is illustrated in Figs.~\ref{Fig_d-wd0001t5}(a-d) and 
\ref{Fig_d-wd0001t5}(e), where the phase transition line $a$ is crossed over on 
increasing $\alpha$ for a given $\lambda$. 

Outside the region with multiple states, the magnitude $q$ of the student 
weight vector decreases above line $c$ (since weight decay becomes strong) or 
to the left of the line $d$ (since examples are not enough). Hence the weight 
vector is not flexible enough to allow for multiple strategies. In general, the 
fraction of sacrificed examples is smaller in this region. This reduces the 
difference between the strategies of sacrificing and not sacrificing the 
examples. As a result, all states to the left of line $b$ learn with a single 
sacrificial strategy and to the right of it with a single even-handed strategy.

\section{CONCLUSION AND REMARKS}
\label{con}

We have studied the supervised learning of of noisy examples in nonlinear and 
differentiable perceptrons using the cavity method. The existence of band gaps 
in the activation distribution is demonstrated and is attributed to the 
competition of conflicting information inherent in noisy data, and the 
nonlinearity of the student perceptron. Activations corresponding to preferred 
or sacrificed examples during learning are separated by band gaps. The band gap 
in the activation distribution is an indication of the extent of information 
competition and the roughness of energy landscape, corresponding to the effects 
of RSB in the replica approach. The more prominent the band gaps, the more 
significant the effects of rough energy landscapes. When 
both the noise and the training set are large, there is a phase transition in 
the student perceptron from a poor generalization state with a long weight 
vector to a good generalization state with a short weight vector. The phase 
transition is accompanied by a change in the learning strategy across the phase 
line from sacrificial to even-handed. We present the phase diagram of this 
system, together with the boundaries of the gapped regime and of the metastable 
region. The relation between band gaps and the picture of a rough energy 
landscape was discussed in a previous study \cite{NIPS}. Here we further show 
where this consideration is most necessary.

We remark that the preferential or sacrificial effects are common in many other 
learning systems, such as multilayer perceptrons \cite{NIPS} and weight pruning 
networks \cite{TANC}. They create metastable states which cause the hysteretic 
behavior as shown in our simulations (see Figs.~9 and 11). The presence of 
metastable states prevent the convergence of dynamical learning process to 
the ground state. Hence it is an important issue in the practical 
implementation of learning dynamics.

We have illustrated that the cavity method can be used to analyze systems laden 
with complex information, yielding predictions identical to the replica method, 
yet providing a more physical interpretation. It can be applied to other 
systems such as Support Vector Machines (SVM) when examples are noisy and 
insufficient \cite{Vapnik}. SVM learning of clean examples has recently been 
studied using the replica theory \cite{Dietrich-op-s}. However, since the 
functional form of the energy is different, band gaps may not be present. 
Nevertheless, a cavity analysis of SVMs could offer new valuable insights. 

\acknowledgments
We thank H. Nishimori and Y. Kabashima for valuable discussions. 
This work is supported by the Research Grant Council of Hong Kong 
(HKUST6157/99P).

\appendix
\section{THE CAVITY ACTIVATION AND LOCAL SUSCEPTIBILITY}
\label{app-a}

From Eqs.~(\ref{equil}) and (\ref{equil_0}) and the definitions of $t_0$ and 
$x_0$, we obtain  
    \begin{equation}			\label{x-t0}    	
    x_0-t_0=\frac{1}{\lambda}(O_0-f_0)f_0'+\frac{1}{\lambda N}\sum_{\mu j}
	[(O_\mu-f_\mu^0)(f_\mu^0)'-(O_\mu-f_\mu)f_\mu']\xi_j^\mu\xi_j^0.
    \end{equation} 
Expanding the last term to first order, and assuming that $x_\mu$ is a well 
defined function of $t_\mu$, we arrive at 
   \begin{eqnarray}			\label{x-t}
   x_0-t_0&=&\frac{1}{\lambda}(O_0-f_0)f_0'+\frac{1}{\lambda N\sqrt{N}}
	    \sum_{\mu j}[-(f_\mu')^2+(O_\mu-f_\mu)f_\mu'']\frac{\partial x_\mu}
	    {\partial t_\mu}\xi_j^\mu\xi_j^0\sum_{k(\neq j)}
             (J_k^{0\backslash\mu}-J_k^{\backslash\mu})\xi_k^\mu  \nonumber \\
	  &&+\frac{1}{\lambda N\sqrt{N}}\sum_{\mu j}[-(f_\mu')^2+
	    (O_\mu-f_\mu)f_\mu'']\frac{\partial x_\mu}{\partial t_\mu}
	   (J_j^{0\backslash\mu}-J_j^{\backslash\mu})(\xi_j^\mu)^2\xi_j^0,  
   \end{eqnarray}
where $J_k^{0\backslash\mu}$ and $J_k^{\backslash\mu}$ denote the student 
weights trained with training sets without example $\mu$, and respectively, 
with and without example 0. Note that 
$\sum_{k(\neq j)} (J_k^{0\backslash\mu}-J_k^{\backslash\mu})\xi_k^\mu/\sqrt{N}
\approx t_\mu^0-t_\mu\sim O(N^{-1/2})$ 
and is uncorrelated with $\xi_j^\mu$. Neglecting the dependence of 
$[-(f_\mu')^2+(O_\mu-f_\mu)f_\mu''](\partial x_\mu/\partial t_\mu)$ 
on $\xi_k^\mu\xi_j^\mu$ that is of order $N^{-1}$, we conclude that the second 
term on the right hand side of (\ref{x-t}) is of order $N^{-1/2}$ 
and hence negligible. In the last term, $(\xi_j^\mu)^2$ is uncorrelated with 
$(J_j^{0\backslash\mu}-J_j^{\backslash\mu})\xi_j^0$, and hence can be replaced 
by its average value of 1. For the remaining summation over $j$, $\sum_j
(J_j^{0\backslash\mu}-J_j^{\backslash\mu})\xi_j^0/\sqrt{N}$ reduces to 
$x_0^{\backslash\mu}-t_0^{\backslash\mu}$. Assuming that the change in the 
activation difference $x-t$ of examples 0 due to the removal of example 
$\mu$ is small, $x_0^{\backslash\mu}-t_0^{\backslash\mu}$ further reduces to 
$x_0-t_0$. Thus
   \begin{eqnarray}			
   x_0-t_0&=&\frac{1}{\lambda}(O_0-f_0)f_0'+\frac{1}{\lambda N}
	\sum_{\mu}[-(f_\mu')^2+(O_\mu-f_\mu)f_\mu'']\frac{\partial x_\mu}
	{\partial t_\mu}(x_0-t_0).
   \end{eqnarray}
Defining the local susceptibility $\gamma$ by
   \begin{equation}						\label{gamma}
   \gamma^{-1}=\lambda+\frac{1}{N}{\sum_{\mu}[(f_\mu')^2-
	(O_\mu-f_\mu)f_\mu'']\frac{\partial x_\mu}{\partial t_\mu}}, 
   \end{equation}
we arrive at Eq.~(\ref{tx}).
Applying the same cavity argument to example $\mu$, $t_\mu$ and $x_\mu$ should 
also be related by Eq.~(\ref{tx}). This simplifies Eq.~(\ref{gamma}) to
   \begin{equation}
   \gamma^{-1}=\lambda+\frac{\gamma^{-1}}{N}\sum_{\mu}\big(\frac{\partial t_\mu}
		{\partial x_\mu}-1\big)\frac{\partial x_\mu}{\partial t_\mu}, 
   \end{equation}
from which Eq.~(\ref{sus}) follows.

\section{THE STABILITY CONDITION}
\label{app-b}

In obtaining Eq.~(\ref{x-t}), the validity of the perturbative expansion in 
Eq.~(\ref{x-t0}) is subject to the condition that the fluctuation 
$\Delta_J=\sum_j(J^0_j-J_j)^2$ is finite. Subtracting Eq.~(\ref{equil_0}) by 
Eq.~(\ref{equil}), multiplying both sides by $J_j^0-J_j$ and summing over $j$, 
we obtain
   \begin{equation}					\label{delta}
   \Delta_J=\frac{1}{\lambda\gamma}(x_0-t_0)^2-\frac{1}{\lambda\gamma}\sum_
      {\mu }\big(1-\frac{\partial x_\mu}{\partial t_\mu}\big)
     \frac{\partial x_\mu}{\partial t_\mu}(t_\mu^0-t_\mu)^2,
   \end{equation}
where Eq.~(\ref{tx}) is adopted, and $x_\mu$ is assumed to be a well defined 
function of $t_\mu$. The factor $(t_\mu^0-t_\mu)^2$ in Eq.~(\ref{delta}) can be 
expanded as $\sum_{jk}(J_j^{0\backslash\mu}-J_j^{\backslash\mu})
(J_k^{0\backslash\mu}-J_k^{\backslash\mu})\xi_j^\mu\xi_k^\mu/N$ 
and is only related with 
$(1-\partial x_\mu/\partial t_\mu)(\partial x_\mu/\partial t_\mu)$ in the order 
$O(N^{-1})$.  
Therefore the average over $\mu$ of the former and later's product  
can be replaced by the product of their averages. Since $(J_j^{0\backslash\mu}
-J_j^{\backslash\mu})(J_k^{0\backslash\mu}-J_k^{\backslash\mu})$ are
uncorrelated with examples ${\bbox \xi}^\mu$, only terms with $j=k$ contribute 
to the average over $\mu$ corresponsively. Then $\sum_\mu(t_\mu^0-t_\mu)^2$ 
becomes $\Delta_J^{\backslash\mu}$. Assuming that the change in $\Delta_J$ due 
to the removal of example $\mu$ is small, this further reduces to $\Delta_J$ 
and renders (\ref{delta}) to
    \begin{eqnarray}				
    \Delta_J&=&\frac{1}{\lambda\gamma}(x_0-t_0)^2-\frac{1}{\lambda\gamma N}
            \sum_\mu\big(1-\frac{\partial x_\mu}
	         {\partial t_\mu}\big)\frac{\partial x_\mu}
	         {\partial t_\mu}\Delta_J.       \nonumber         
    \end{eqnarray}
Using the relation (\ref{tx}) between the generic and cavity activations 
of example $\mu$, this can be further reduced to Eq.~(\ref{Delta_J}).

\section{THE EFFECTS OF A GAP ON MEAN FIELD EQUATIONS}
\label{app-c}

When there is a gap in the distribution $P(x|{\tilde y})$ of student activation 
$x$ for a given teacher output ${\tilde y}$, the mean field equations 
(\ref{susi2}), (\ref{R}) and (\ref{q}) are not exact since $x_\mu$ is no longer 
a differentiable function of $t_\mu$. Nevertheless, we can obtain its modified 
expression from self-consistent considerations. 

In Eq.~(\ref{x-t0}), the summation over $\mu$ now includes different situations 
depending on the value of the cavity field $t_\mu$. For those examples with 
$t_\mu$ and $t_\mu^0$ located on the same side of the gap, the analysis is 
similar to the that in Appendix A. However, if $t_\mu$ is close to the gap
position $t_g$, then when the new example 0 is included in the training set, 
the change of cavity activation $\Delta t_\mu\equiv\sum_k(J_k^{0\backslash\mu}
-J_k^{\backslash\mu})\xi_k^\mu/{\sqrt N}$ may give rise to large value of 
$(O_\mu-f_\mu^0)(f_\mu^0)'-(O_\mu-f_\mu)f_\mu'$ 
as the generic activation $x_\mu$ changes from $x_<(t_\mu)$ to above 
$x_>(t_\mu)$ or reverse. We distinguish the following cases to calculate the 
summation in Eq.~(\ref{x-t0}).

The first case corresponds to $t_g-\Delta t_\mu<t_\mu<t_g$. Among the $p$ 
examples, this happens with probability $\delta{\bbox (}t_\mu
-t_g({\tilde y_\mu}){\bbox )}\Delta t_\mu\theta(\Delta t_\mu)$. 
Its contribution to the summation in Eq.~(\ref{x-t0}) is 
    \begin{eqnarray}
    \sum_{\{{\rm Case\,1}\}}&=&\frac{1}{\lambda N}\sum_{\mu j}
        \delta{\bbox (}t_\mu-t_g({\tilde y_\mu}){\bbox )}\Delta t_\mu
         \theta(\Delta t_\mu)					\nonumber \\ 
     &&\times\{[f({\tilde y_\mu})-f(x_\mu^>)]f'(x_\mu^>)
          -[f({\tilde y_\mu})-f(x_\mu^<)]f'(x_\mu^<)\}\xi_j^\mu\xi_j^0.  
    \end{eqnarray} 
Similarly, the second case corresponds to $t_g<t_\mu<t_g-\Delta t_\mu$, with
the contribution   
    \begin{eqnarray}
    \sum_{\{{\mathrm Case\,2}\}}&=&\frac{1}{\lambda N}\sum_{\mu j}\delta
       {\bbox (}t_\mu-t_g({\tilde y_\mu}){\bbox )}(-\Delta t_\mu)
               \theta(-\Delta t_\mu)  				\nonumber \\
       &&\times\{[f({\tilde y_\mu})-f(x_\mu^<)]f'(x_\mu^<)
         -[f({\tilde y_\mu})-f(x_\mu^>)]f'(x_\mu^>)\}\xi_j^\mu\xi_j^0.  
    \end{eqnarray} 
Combining them together, we have the total contribution from the gap
    \begin{eqnarray}
    \sum _{\{{\rm Gap}\}}&=&\frac{\alpha}{\lambda}(x_0-t_0) \int{\rm d}yP(y)
       \int{\rm d}{\tilde y}P({\tilde y}|y)\int{\rm d}tP(t|{\tilde y})
       \delta{\bbox (}t-t_g({\tilde y}){\bbox )}             \nonumber \\
       &&\times{\bbox \{}{\bbox [}f({\tilde y})-f{\bbox (}x_>(t,{\tilde y})
       {\bbox )}{\bbox ]}f'{\bbox (}x_>(t,{\tilde y}){\bbox )}-{\bbox [}
       f({\tilde y})-f{\bbox (}x_<(t,{\tilde y}){\bbox)}{\bbox ]}
       f'{\bbox (}x_<(t,{\tilde y}){\bbox )}{\bbox \}}.
    \end{eqnarray}
Simplifying the integrals, we have
    \begin{eqnarray} 
    \sum_{\{{\rm Gap}\}}&=&\frac{\alpha}{\lambda}\frac{x_0-t_0}
      {\sqrt{2\pi\big(q-\frac{R^2}{1+T^2}\big)}}\int{\rm D}u\,
      \exp\Big[-\frac{\big(t_g-\frac{R}{\sqrt{1+T^2}}u\big)^2}
      {2\big(q-\frac{R^2}{1+T^2}\big)}\Big]		\nonumber	\\
       &&\times{\bbox \{}{\bbox [}f({\tilde y})-f{\bbox (}x_>(t_g,{\tilde y})
       {\bbox )}{\bbox ]}f'{\bbox (}x_>(t_g,{\tilde y}){\bbox )}-{\bbox [}
       f({\tilde y})-f{\bbox (}x_<(t_g,{\tilde y}){\bbox)}{\bbox ]}
       f'{\bbox (}x_<(t_g,{\tilde y}){\bbox )}{\bbox \}},
    \end{eqnarray}
with ${\tilde y}=\sqrt{1+T^2}u$. Therefore, we obtain the self-consistent 
equation (\ref{sus_e}) for $\gamma$ and the function $t(x)$ in Eq.~(\ref{tx}), 
where $x$ is related to $u$ and $v$ by Eq.~(\ref{xuv}). The positions of band 
gap $t_g$, $x_<$ and $x_>$ are determined using the Maxwell's construction 
discussed in Sec.~\ref{bandgap}. Following Eqs.~(\ref{sus_e}), (\ref{tx}) and 
(\ref{xuv}), we get the equation of $R$ with extra terms (\ref{R_e}) and 
equation of $q$ without extra term (\ref{q}), after elaborate work on 
integrating by parts.

\section{CONDITION FOR MAXWELL'S CONSTRUCTION}

For a given teacher output $f({\tilde y})$, $x$ is a multi-valued function of
$t$ when $t'(x)<0$ at the inflection point $t''(x)=0$. For the sigmoid function 
$f(x)=[1+e^{-x}]^{-1}$, this implies 
   \begin{equation}					\label{f-deri}
   \frac{1}{2\gamma}<\frac{f^2(1-f)^2(1-3f+3f^2)}{1-6f+6f^2},
   \end{equation}   
where $f$ represents $f(x)$ at the inflection point and can be solved by
   \begin{equation}					\label{s-deri}
   f({\tilde y})=\frac{f(4-15f+12f^2)}{1-6f+6f^2}.
   \end{equation}
Note that the conditions (\ref{f-deri}) and (\ref{s-deri}) are invariant
when $f$ and $f({\tilde y})$ transform to $1-f$ and $1-f({\tilde y})$ 
respectively. Hence the region of Maxwell's construction are symmetric with 
respect to the point ${\bbox (}f,f({\tilde y}){\bbox )}=(1/2,1/2)$. Thus, we 
obtain the condition of Maxwell's construction (\ref{W_ga}) if we define $W$ as 
the parametric function of $\gamma$ via 
   \begin{equation}					 \label{W_f}   
   W=\frac{(2\tilde{f}-1)(1-12\tilde{f}+12\tilde{f}^2)}
		{2(1-6\tilde{f}+6\tilde{f}^2)}    
   \end{equation}
and
   \begin{equation}     				 \label{gamma_f}
   \gamma=\frac{1-6\tilde{f}+6\tilde{f}^2}
    {2\tilde{f}^2(1-\tilde{f})^2(1-3\tilde{f}+3\tilde{f}^2)}.  
   \end{equation}						
The function $W(\gamma)$ for $\gamma>0$ is plotted in 
Fig.~\ref{Fig_W-gamma}.


\begin{figure} 
\caption{
The Maxwell's construction to determine the position of band gap $t_g$. In the 
figure, the areas of the two shaded regions equal to each other. The labels of 
the points are to be compared with the lines in Fig.~\ref{Fig_Maxwell}.
}
\label{Fig_x-t}
\end{figure}

\begin{figure}
\caption{
The function $W$ of $\gamma$ defined by Eq.~(\ref{W_f}) and 
Eq.~(\ref{gamma_f}), which is introduced to determine the condition of band gap 
occurrence.
}
\label{Fig_W-gamma}
\end{figure}

\begin{figure} 
\caption{
The occurrence of band gap and preferential learning when $\gamma=20.55$. The
picture is symmetric with respect to the point (1/2,1/2). For intermediate 
teacher output $0.136<f({\tilde y})<0.864$, no band gaps exist. 
For related discussions in Figs.~\ref{Fig_gap-regime}, \ref{Fig_d-a3wd0002}
and Table \ref{table1} when $\alpha=3$ and $\lambda=0.002$, the present value 
of $\gamma$ corresponds to $T=2.5$, and the choice of ${\tilde y}=-\sqrt{1+T^2}$
corresponds to the line of $f({\tilde y})=0.063$, which cuts the boundaries of 
the shaded region at $x=1.18$ and $x=2.78$, indicating a band gap of 
$P(x|{\tilde y})$ at [1.18,2.78]. 
}
\label{Fig_Maxwell}
\end{figure}

\begin{figure} 
\caption{
Regimes of the existence of gapped activation distribution and the regimes 
of unstable states for different noise levels. Below the solid lines, the 
perturbative cavity solutions are unstable, and below the dotted line, band
gaps will appear in the activation distribution. The point 
$(\alpha=3,\lambda=0.002)$ is denoted by a star. The shaded regions indicate 
the existence of discontinuous phase transitions to be discussed in 
Sec.~\ref{phase}. (For $T=0.1$, the shaded region is too small to be shown.) 
}
\label{Fig_gap-regime}
\end{figure}

\begin{figure} 
\caption{
Theoretical and simulation results of student activation distributions, 
indicated by solid and dashed lines respectively, when  
$\alpha=3$, $\lambda=0.002$ (denoted by a star in Fig.~(\ref{Fig_gap-regime})) 
and $\tilde{y}=-\sqrt{1+T^2}$ for different noise temperatures. 
The arrow in (b) shows the position of a pseudo gap and the arrows in (c) show 
the band gap [1.18, 2.78] from the theoretical prediction in 
Fig.~(\ref{Fig_Maxwell}).
}
\label{Fig_d-a3wd0002}
\end{figure}

\begin{figure} 
\caption{
The theoretical prediction of the student activation distributions at $T=5$ 
and $\lambda=0.001$ for different sizes of the training set $\alpha$, where 
$\tilde{y}=0$. When $\alpha=5$, the system has two states. Respectively,
(d) and (e) are the distributions when the system is in the poor and good 
generalization states.
}
\label{Fig_d-wd0001t5}
\end{figure}

\begin{figure} 
\caption{
The dependence of local susceptibility $\gamma$ on the weight decay strength 
$\lambda$ for different $\alpha$ at $T=1$. All curves approach $\lambda=0$ 
when $\gamma$ goes to infinity.
}
\label{Fig_lambda-ga}
\end{figure}

\begin{figure} 
\caption{
The dependence of generalization error $\varepsilon_g$ on the weight decay 
strength $\lambda$ for different sizes $\alpha$ of training set when the noise 
temperature $T=1$. Inset illustrates the three branches of energy curve for 
$\alpha=1.9$.
}
\label{Fig_energy-eg-t1}
\end{figure}

\begin{figure} 
\caption{
Variations of $\varepsilon_g$, $\varepsilon_t$, $R$ and $q$ of a sample of 
simulation when the weight decay strength $\lambda$ changes at $T=5$, 
$\alpha=4$ and $N=50$. The arrows in (a) denote the routes of changing.
}
\label{Fig_hestera4t5}
\end{figure}

\begin{figure}
\caption{
The learning curves for different weight decay strengths: (a) $T=1$; (b) $T=5$. 
}
\label{Fig_t1-t5-wd}
\end{figure}

\begin{figure} 
\caption{
Simulation versus theoretical results for the generalization error on 
changing $\alpha$. (a) T=1 and $\lambda=0.0001$, the simulation result is the  
average over 14 samples. 
(b) $T=5$ and $\lambda=0.001$, the simulation result is the average over 20 
samples on decreasing and increasing $\alpha$. In all simulations,
the number of input nodes $N=150$. 
}
\label{Fig_t1-t5-eg}
\end{figure}

\begin{figure} 
\caption{
The energy $E/N$ (solid line) and the magnitude of student vector $q$ (dotted 
line) versus the size $\alpha$ of training set, $T=5$ and $\lambda=0.001$. The 
phase transition point is determined from the crossover of the two branches of 
the energy curve, $\alpha_c^t=5.95$, and the spinodal point of the good 
generalization state is at $\alpha_g^{sp}=4.4$.   
}
\label{Fig_t5-e-q-a} 
\end{figure}

\begin{figure} 
\caption{
The phase diagram for nonlinear perceptrons learning noisy examples when $T=1$.
$P$ is the critical point with $\alpha=\alpha_c^*=1.65$. Line $d$ terminates
at $\alpha=\alpha_0=1.737$ when $\lambda$ approaches zero.
}
\label{Fig_phase-diagram-t1} 
\end{figure}

\begin{table}
\caption{The comparison of macroscopic parameters and errors obtained from 
theory (Roman) and simulation (italics in brackets) for different $T$ when 
$\alpha=3$ and $\lambda=0.002$.}  
\label{table1}
\begin{tabular}{cccccc}                               
$T$   & $\gamma$    &$R$   &$q$   &$\epsilon_t$    &$\epsilon_g$  \\\hline
0.1  & 11.1 & 0.963 ({\it0.961})  & 0.933 ({\it0.932})  & 0.018 ({\it 0.017})
							& 0.027 ({\it 0.027})\\
2.0  & 14.9 & 0.796 ({\it 0.795}) & 2.211 ({\it 2.196}) & 0.236 ({\it 0.236})
						        & 0.376 ({\it 0.375})\\
2.5  & 20.6 & 0.837 ({\it 0.822}) & 3.816 ({\it 3.574}) & 0.260 ({\it 0.260})
						        & 0.437 ({\it 0.431})\\
5.0  & 80.1 & 0.920 ({\it 0.848}) & 23.05 ({\it 16.34}) & 0.275 ({\it 0.301})
						        & 0.577 ({\it 0.570})\\
\end{tabular}    
\end{table}

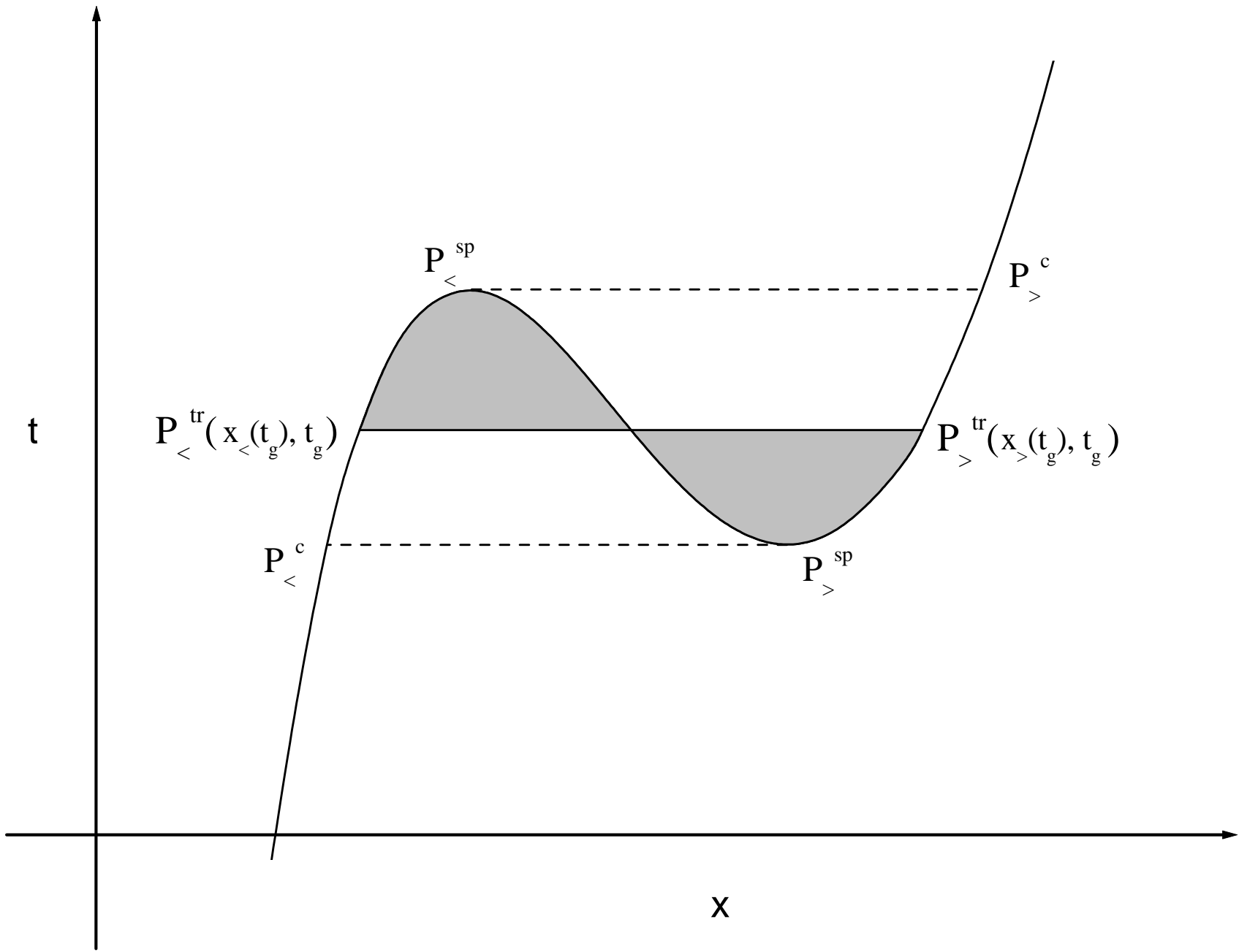
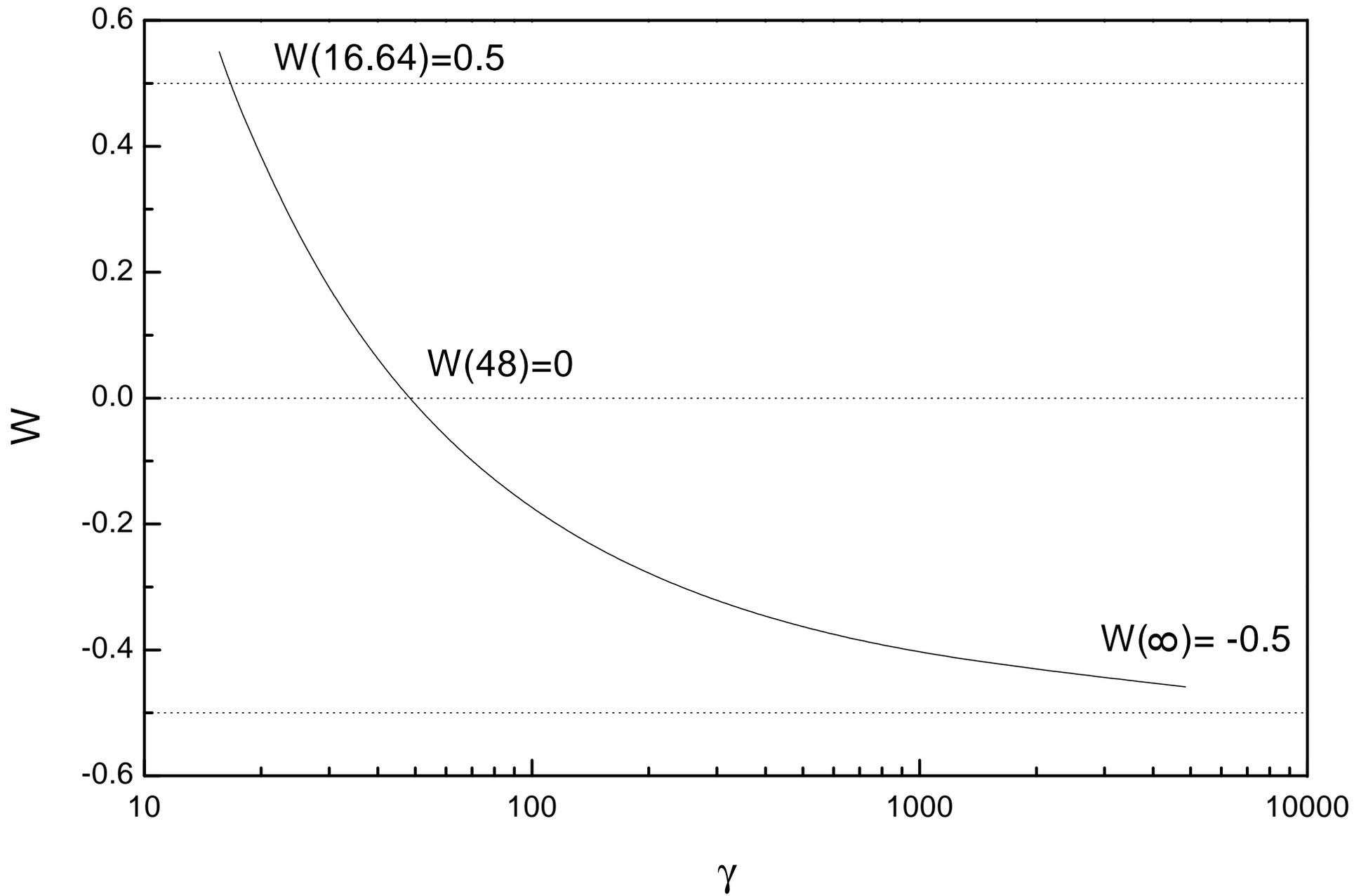
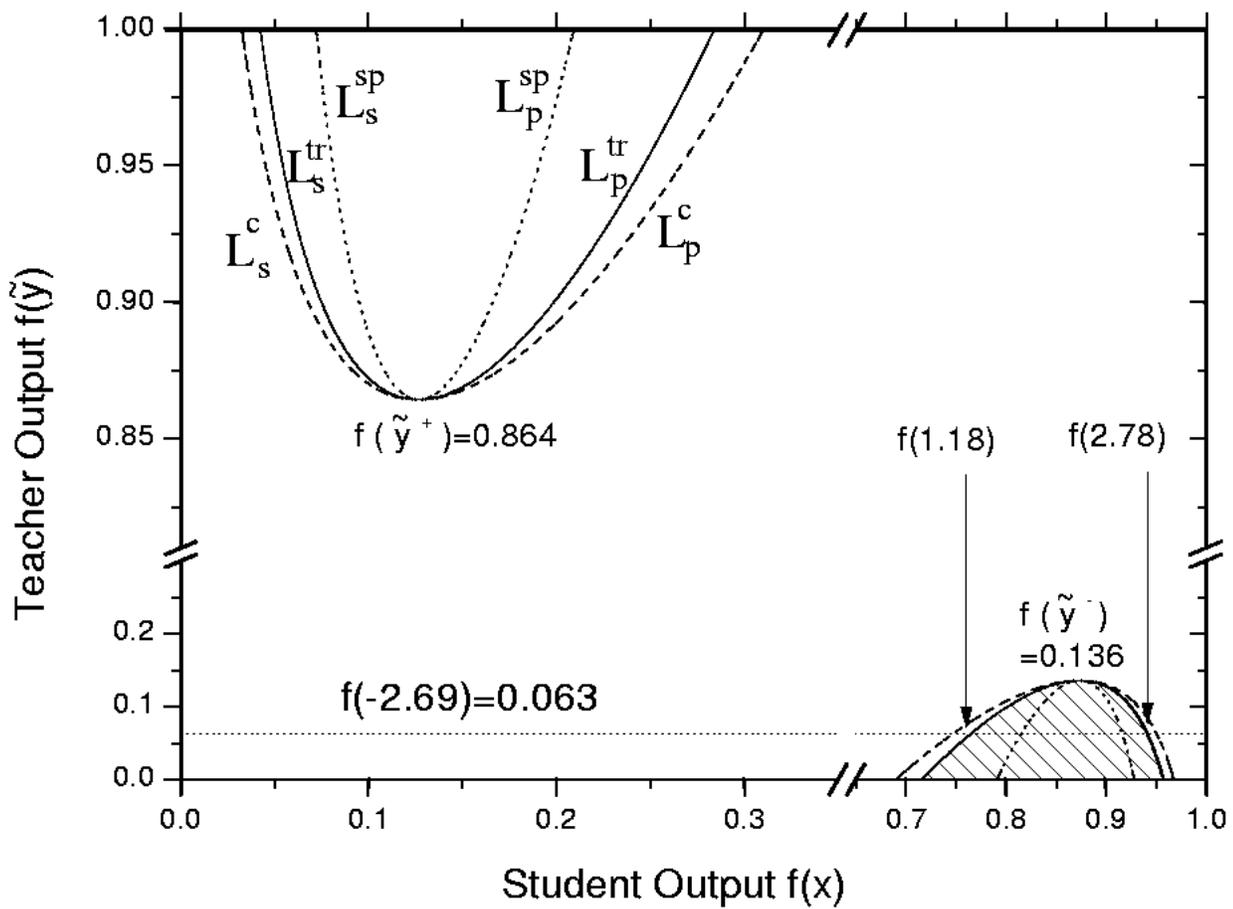
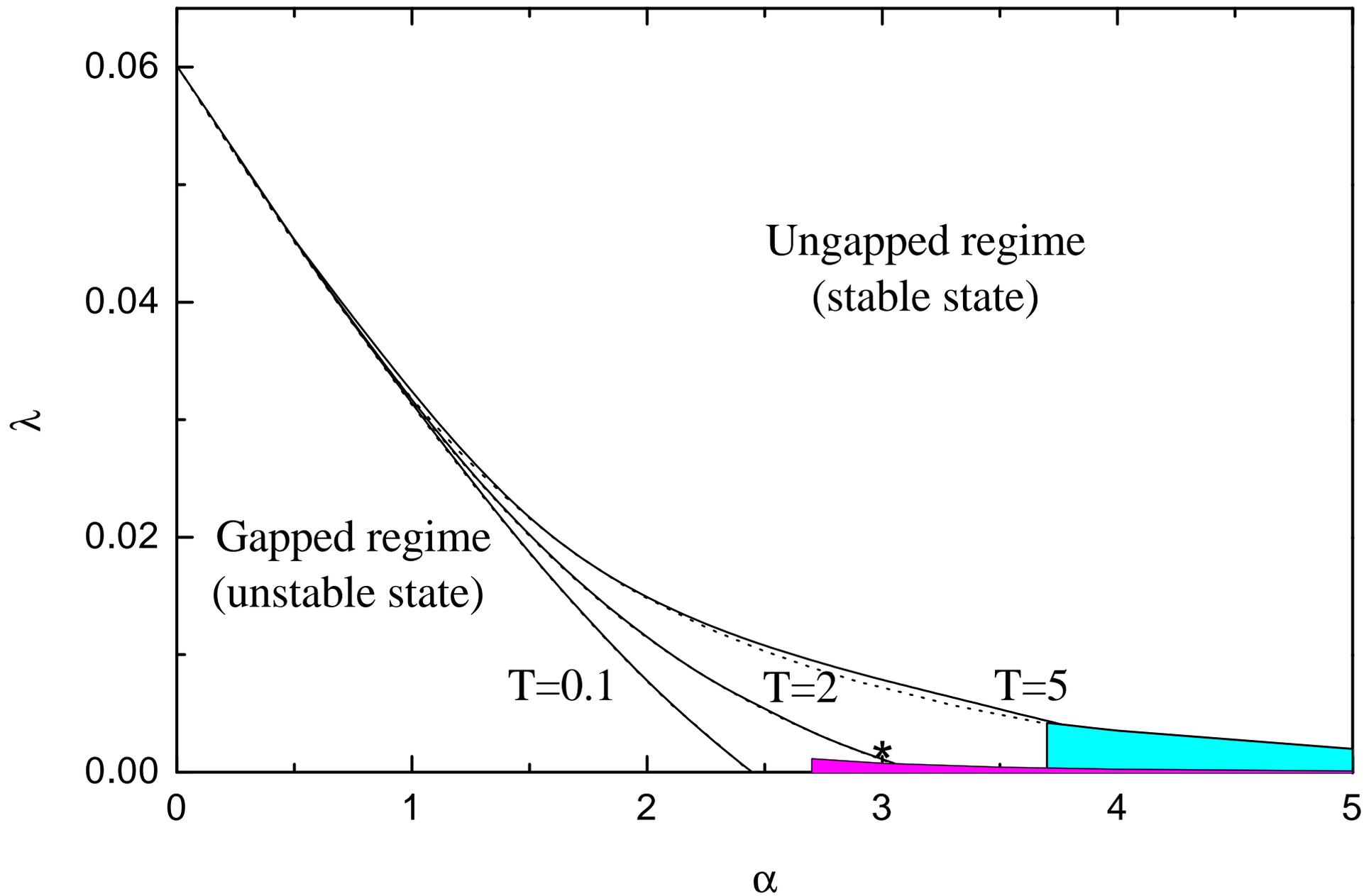
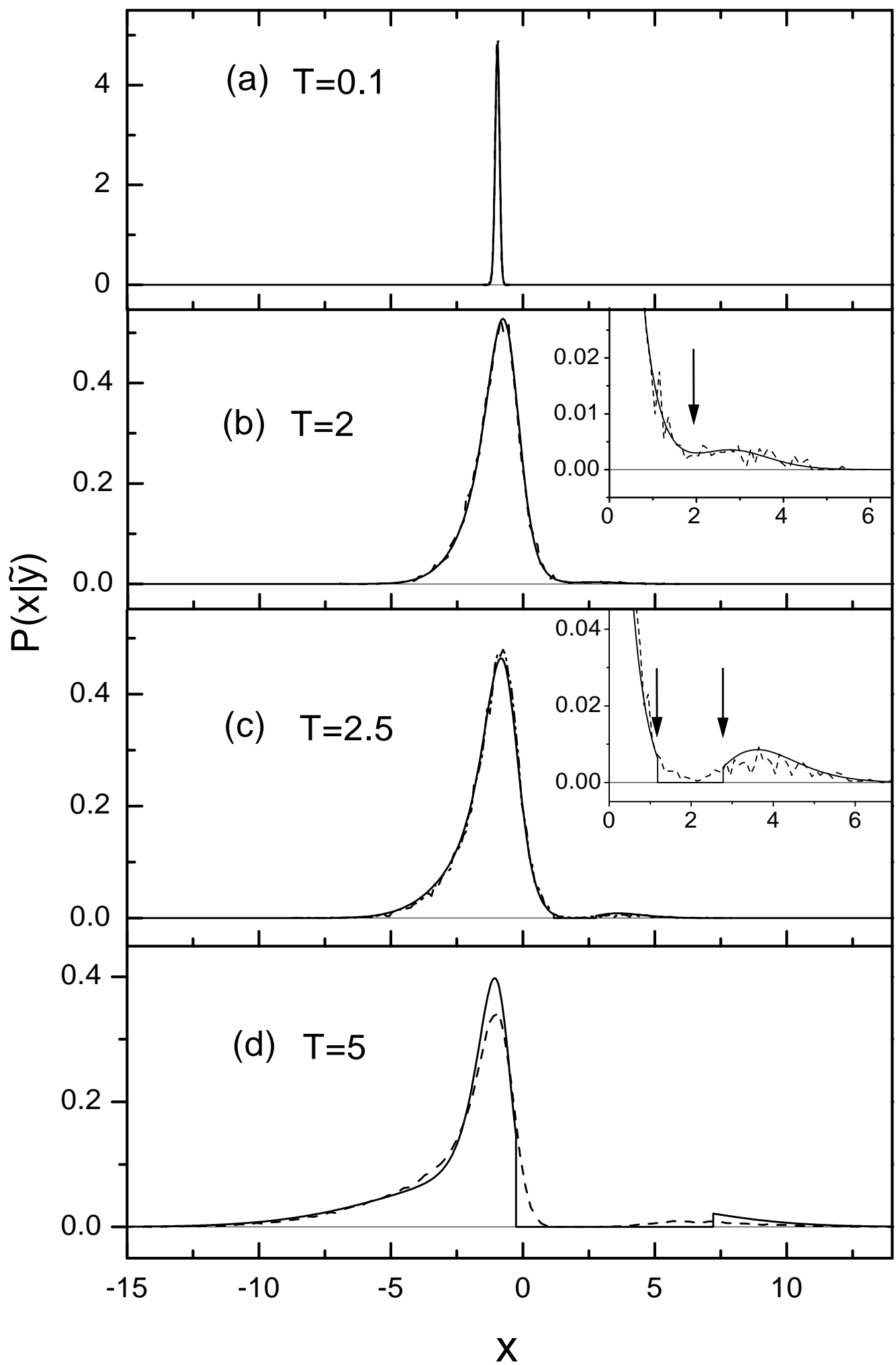
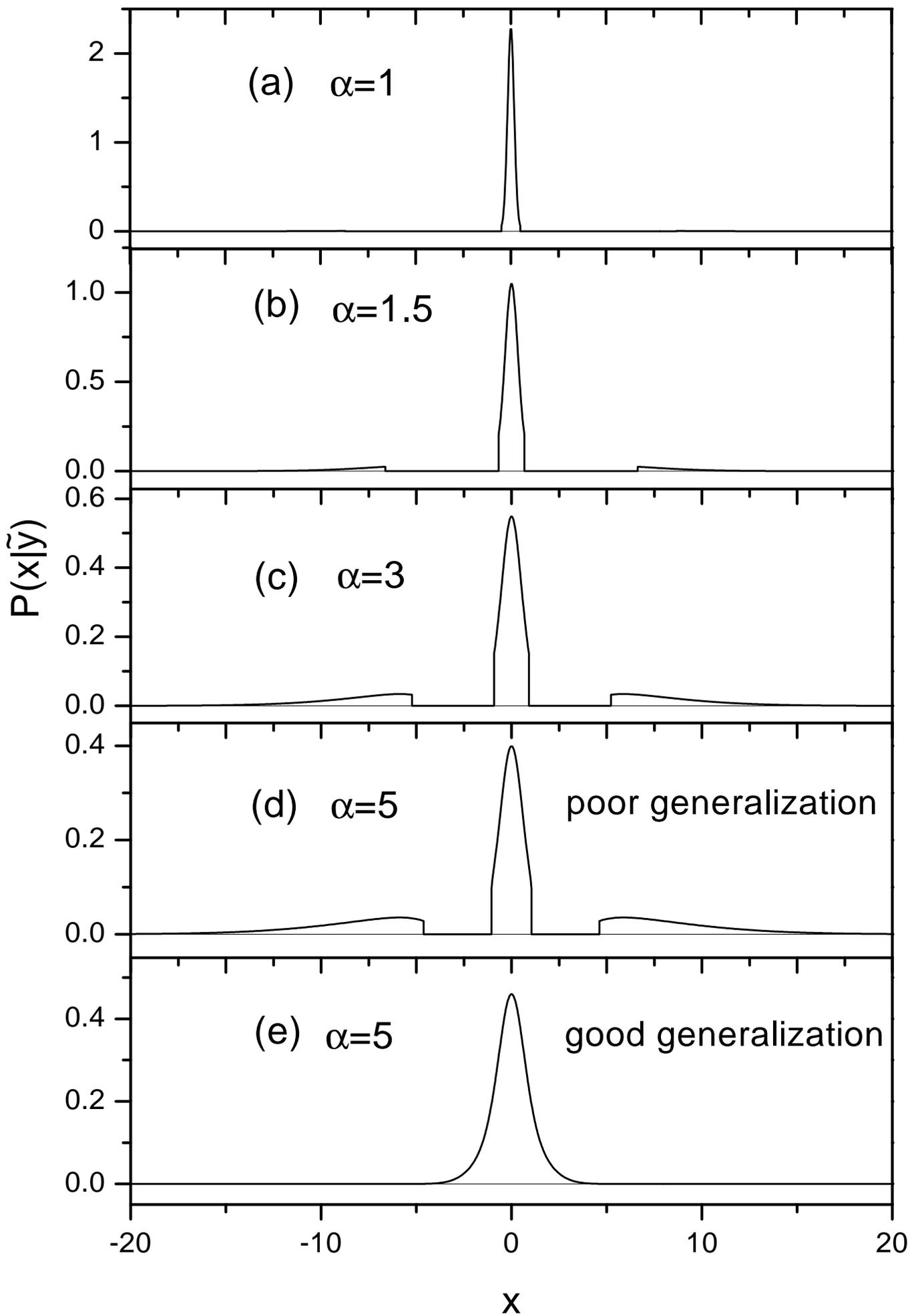
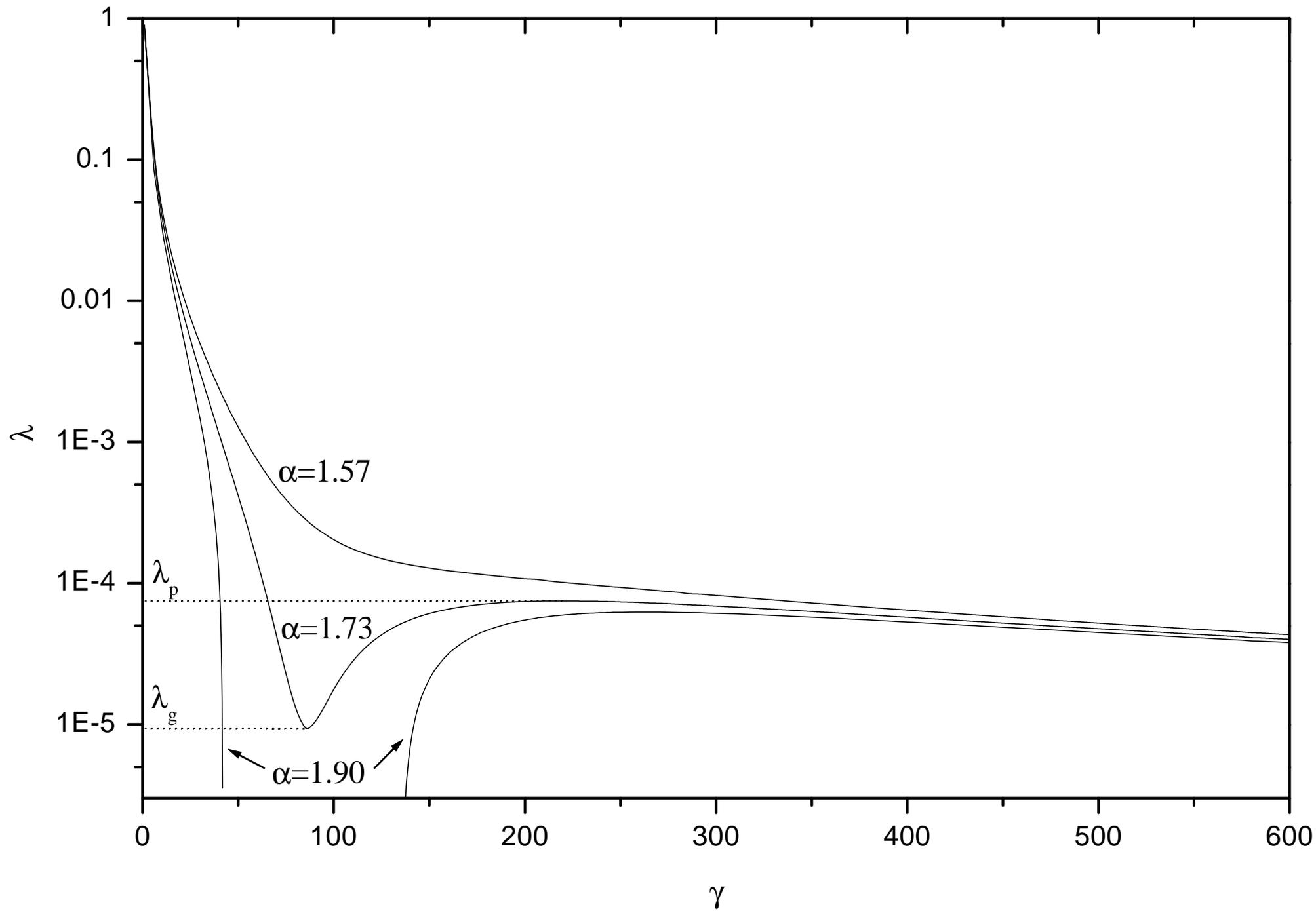
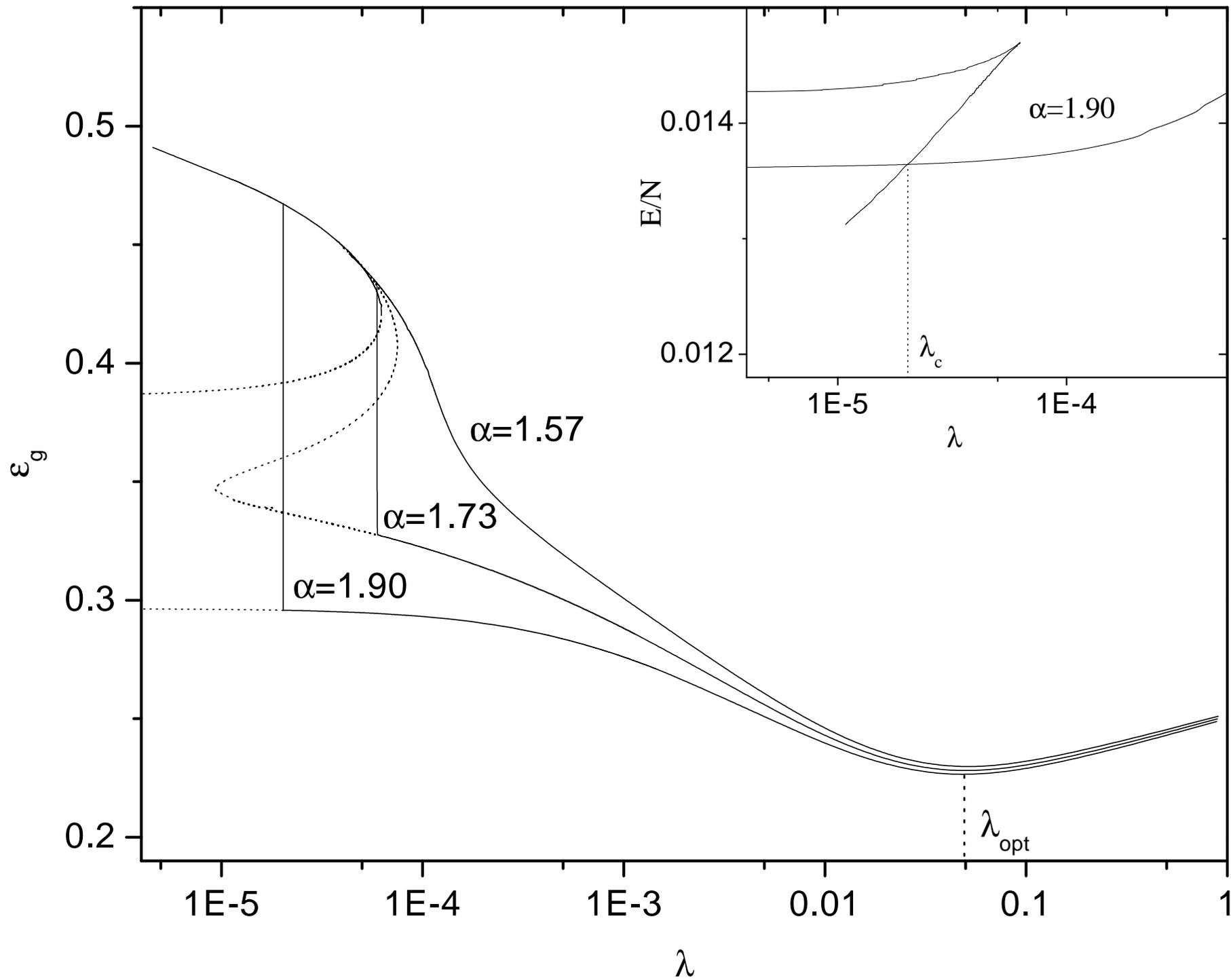
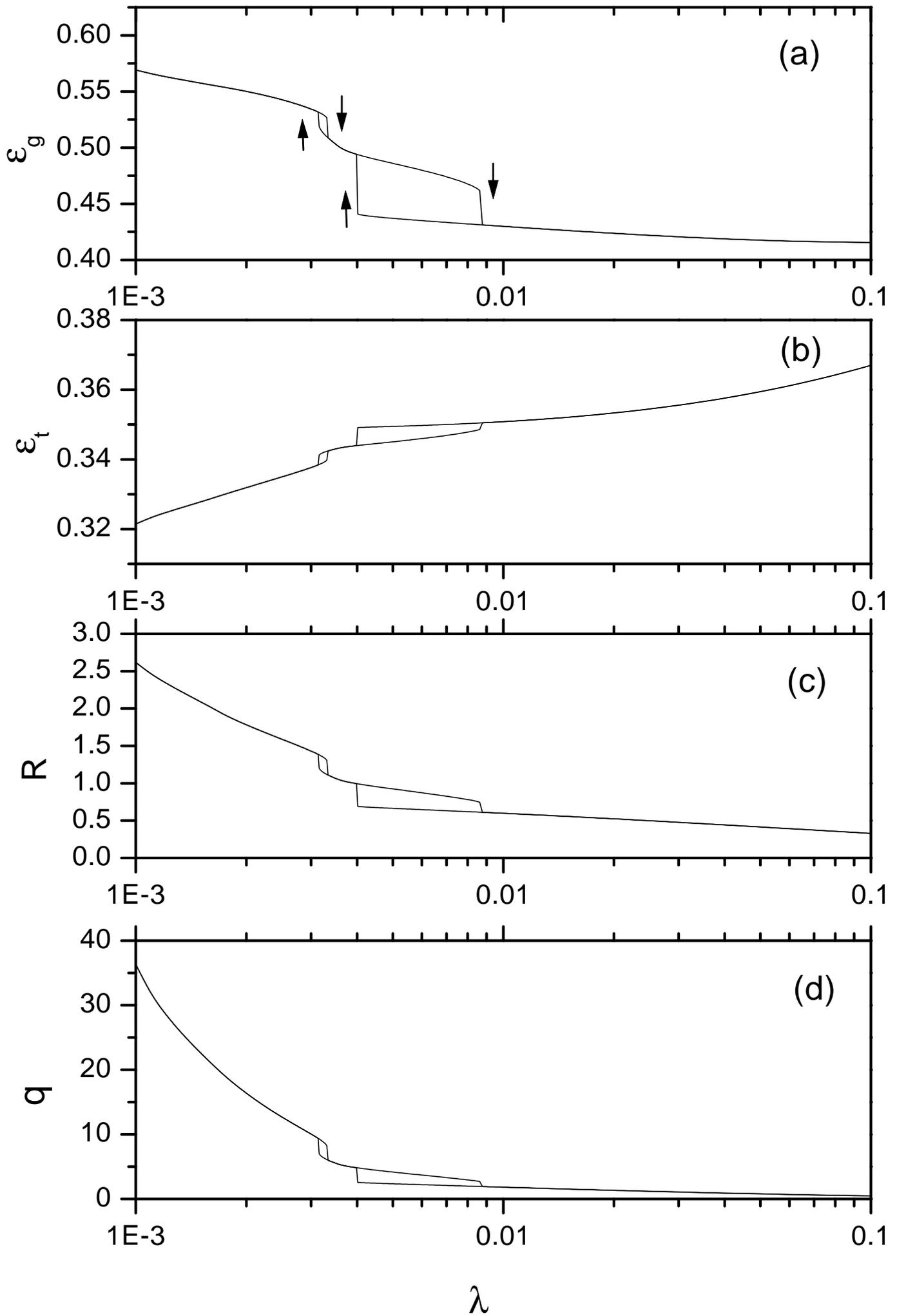
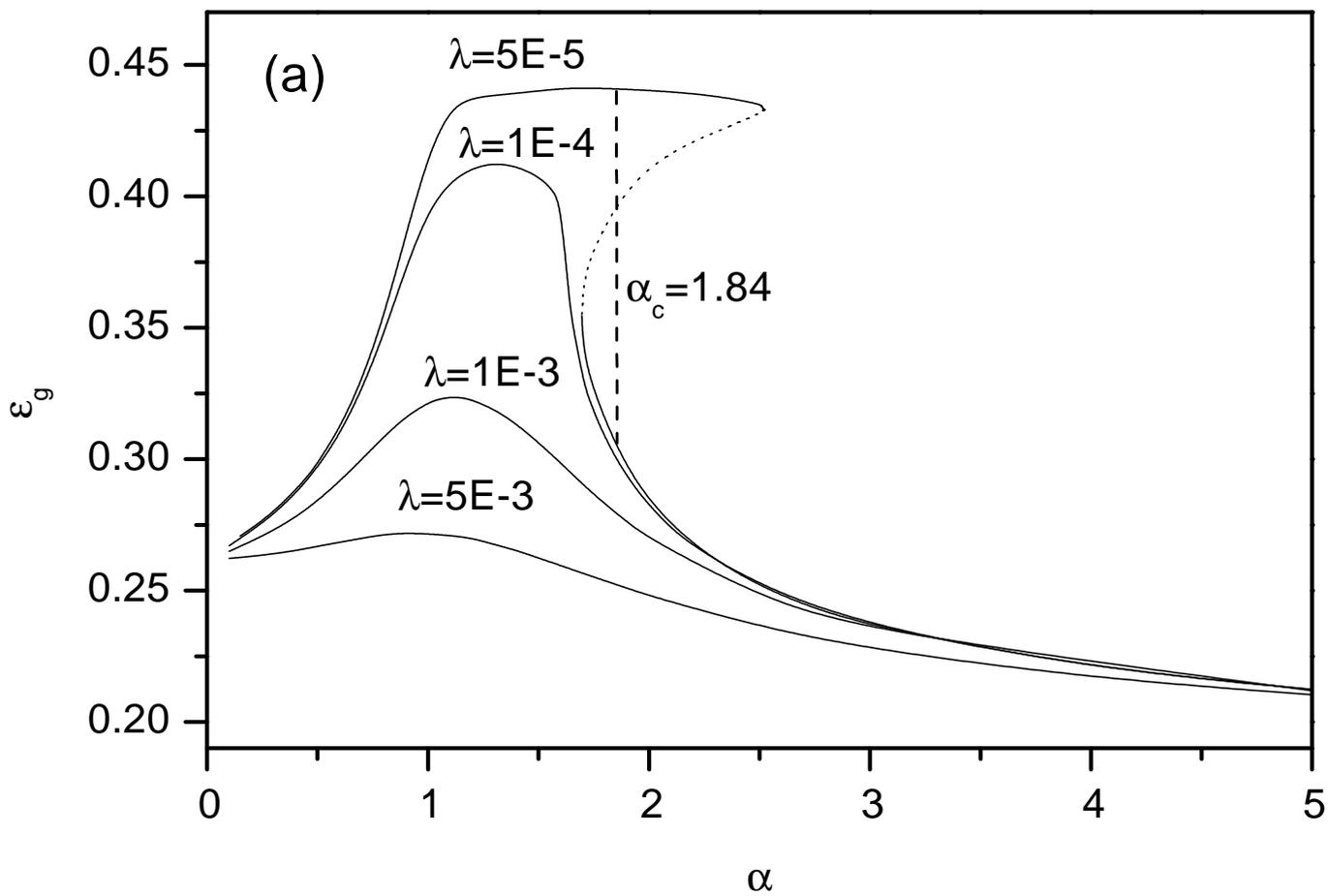
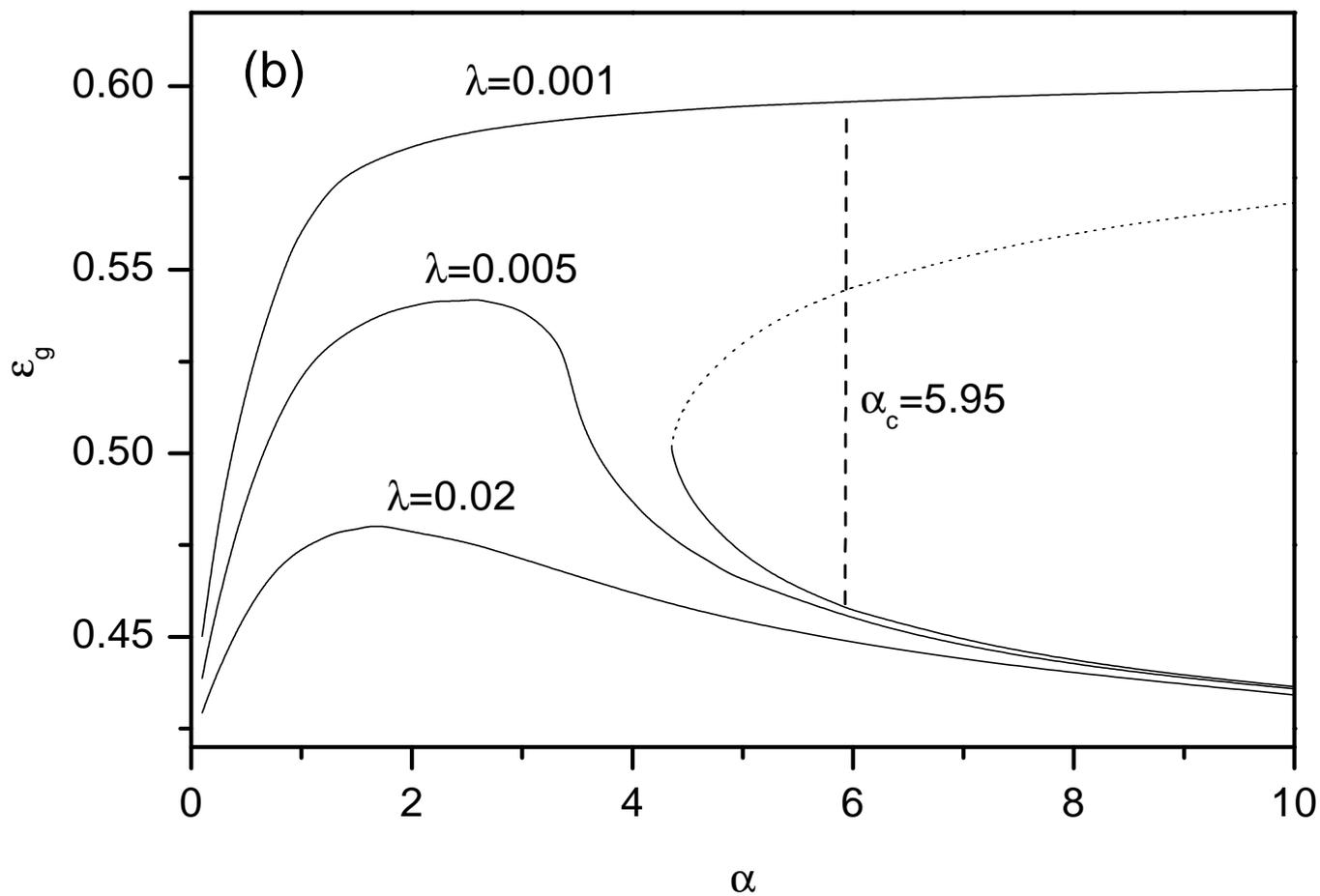
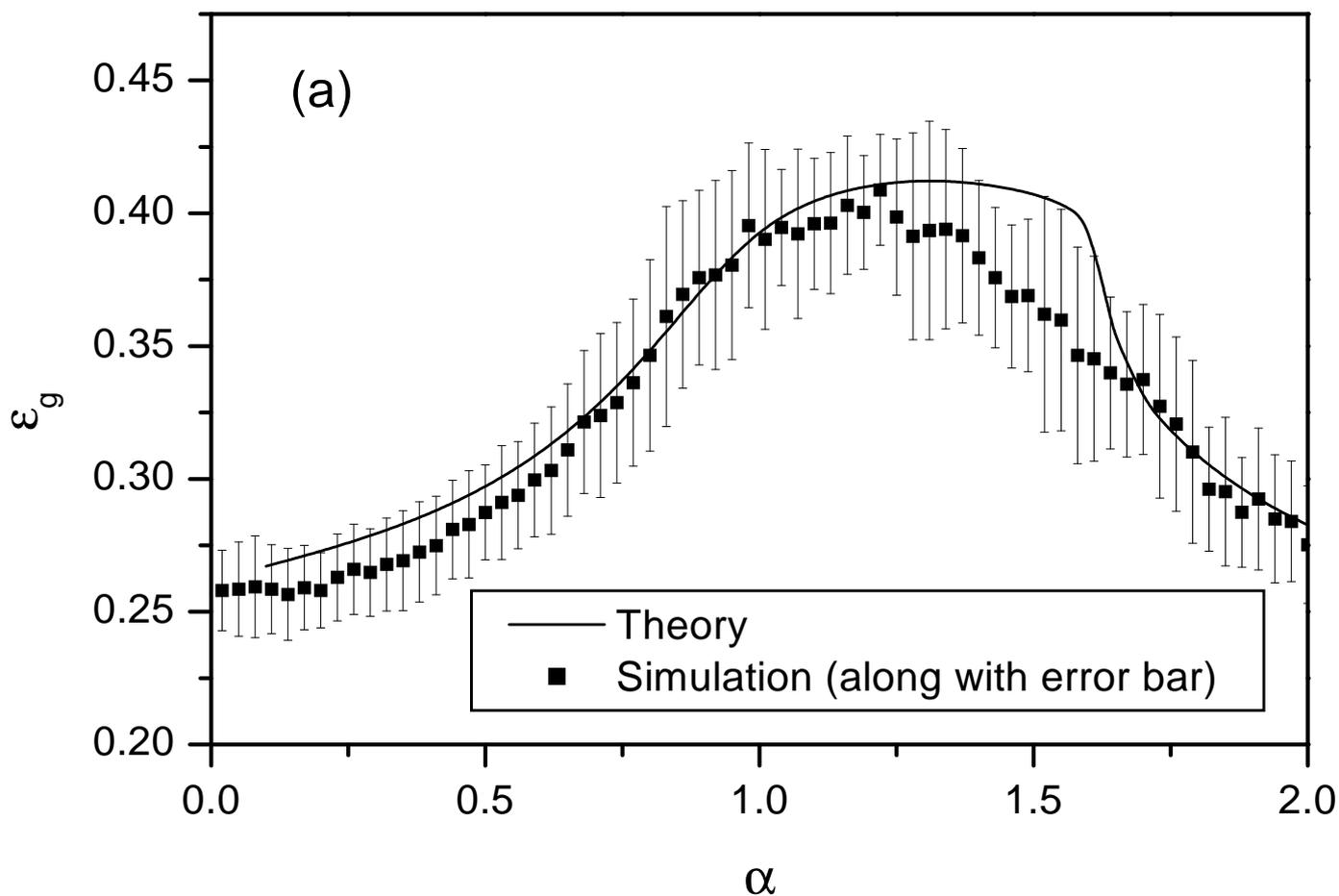
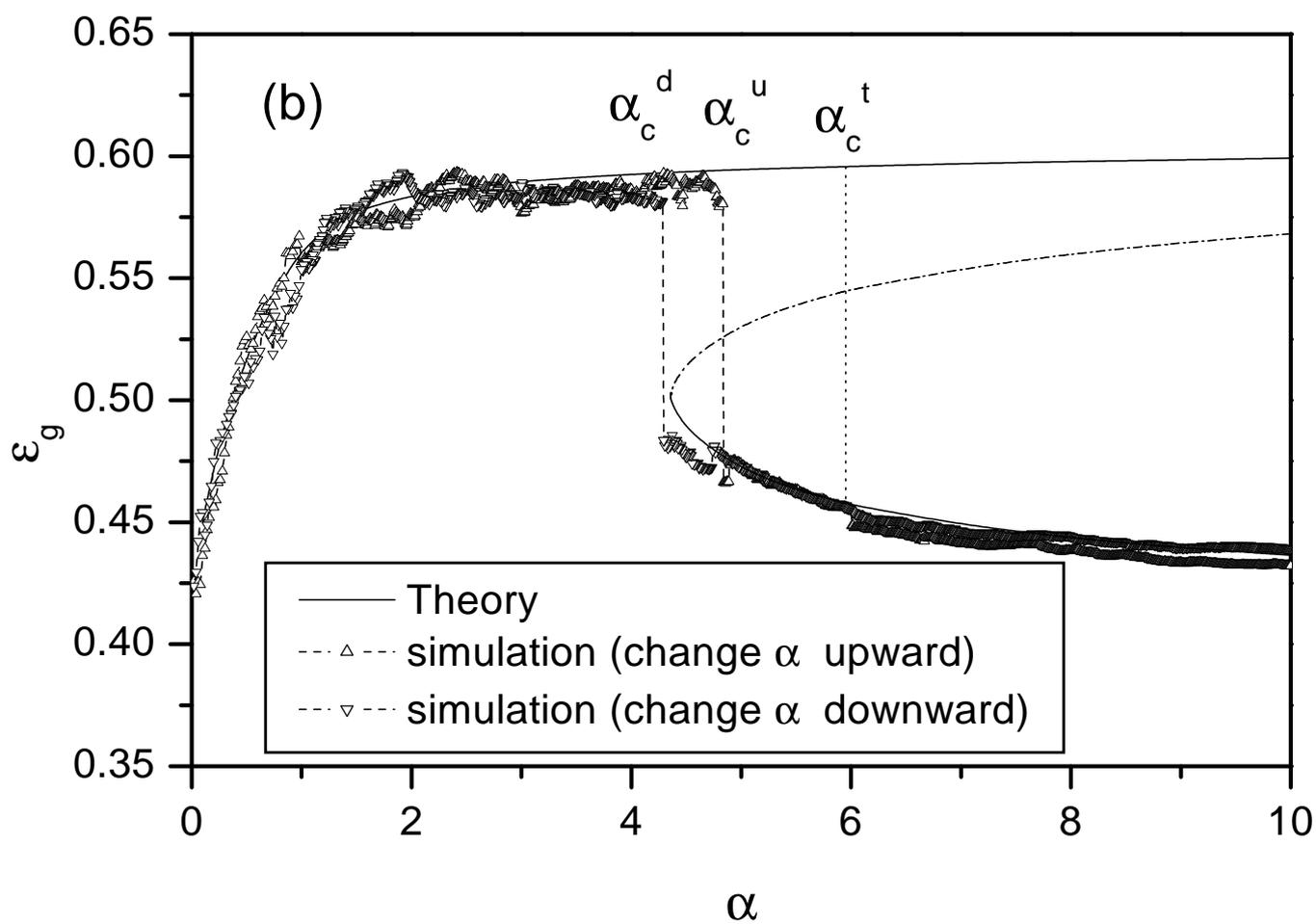
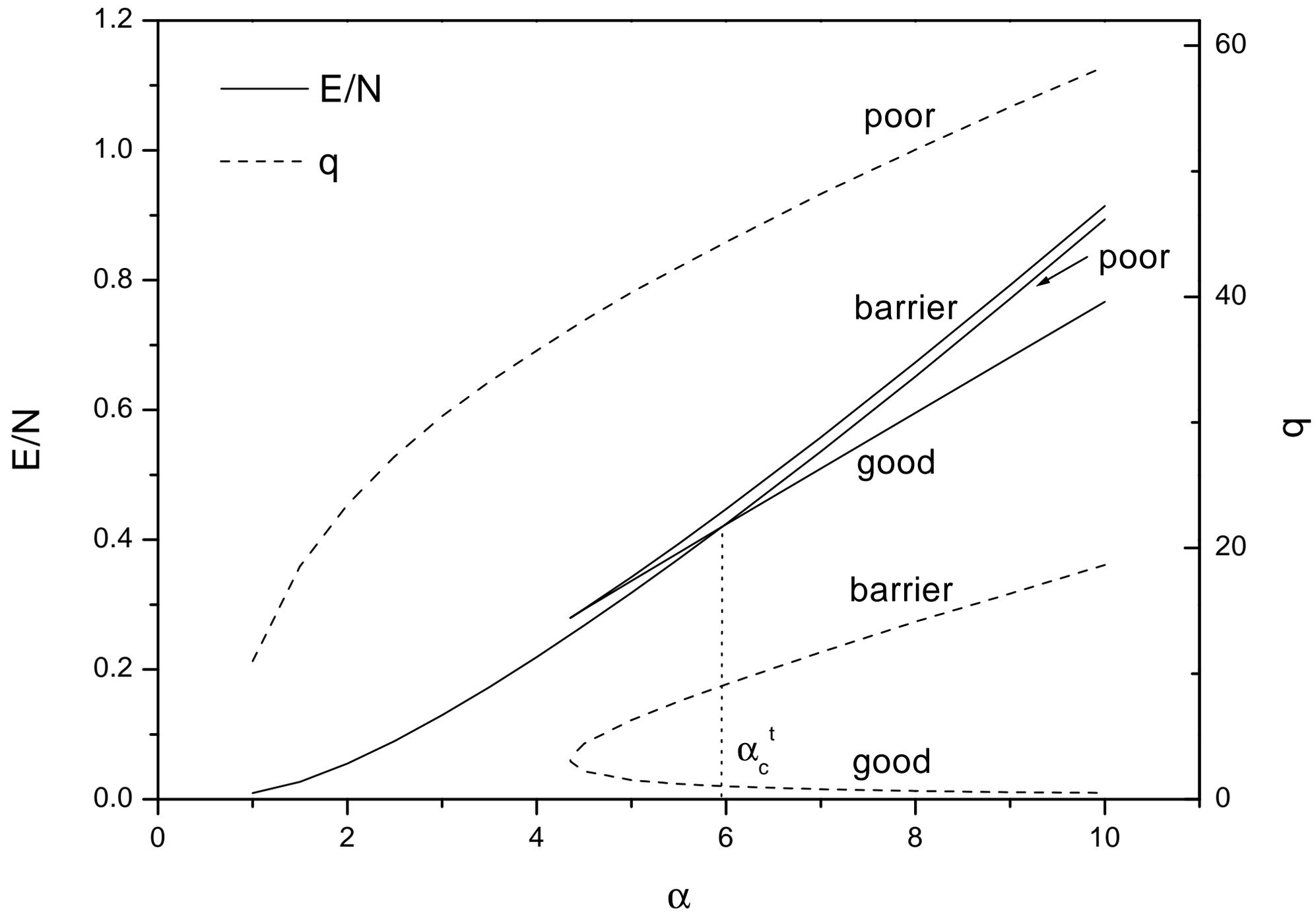
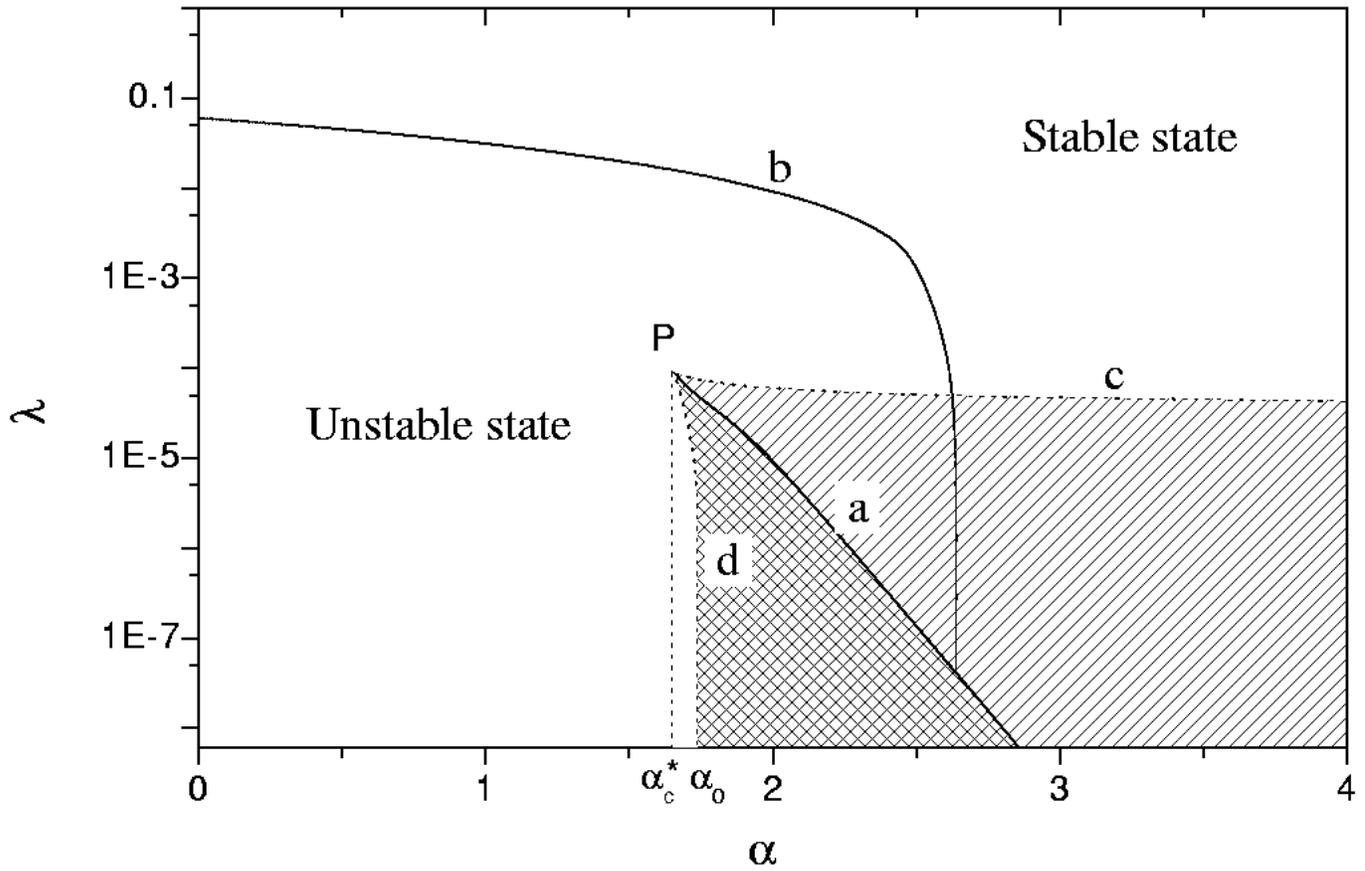
\end{document}